\documentclass[]{aastex631}
\usepackage{CJK}

\begin{document}
\begin{CJK*}{UTF8}{ipxm}

\title{GR-RMHD Simulations of Super-Eddington Accretion Flows onto a Neutron Star with Dipole and Quadrupole Magnetic Fields}

\author[0000-0002-0700-2223]{Akihiro Inoue}
\affiliation{Department of Earth and Space Science, Graduate School of Science, Osaka University, Toyonaka, Osaka 560-0043, Japan}
\email{ainoue@astro-osaka.jp}

\author[0000-0002-2309-3639]{Ken Ohsuga}
\affiliation{Center for Computational Sciences, University of Tsukuba, 1-1-1 Ten-nodai, Tsukuba, Ibaraki 305-8577, Japan}

\author[0000-0003-0114-5378]{Hiroyuki R. Takahashi}
\affiliation{Department of Natural Sciences, Faculty of Arts and Sciences, Komazawa University, Tokyo 154-8525, Japan}

\author[0000-0003-3640-1749]{Yuta Asahina}
\affiliation{Center for Computational Sciences, University of Tsukuba, 1-1-1 Ten-nodai, Tsukuba, Ibaraki 305-8577, Japan}

\author{Matthew J. Middleton}
\affiliation{School of Physics \& Astronomy, University of Southampton, Southampton, Southampton SO17 1BJ, UK}



\begin{abstract}

Although ultraluminous X-ray pulsars (ULXPs) are believed to 
be powered by super-Eddington accretion onto a magnetized neutron star (NS), 
the detailed structures of the inflow-outflow and magnetic fields 
are still not well understood. 
We perform general relativistic radiation magnetohydrodynamics (GR-RMHD) 
simulations of super-Eddington accretion flows 
onto a magnetized NS with dipole and/or quadrupole magnetic fields. 
Our results show that an accretion disk and optically thick outflows form 
outside the magnetospheric radius, 
while inflows aligned with magnetic field lines appear inside. 
When the dipole field is more prominent 
than the quadrupole field at the magnetospheric radius, 
accretion columns form near the magnetic poles, 
whereas a quadrupole magnetic field stronger 
than the dipole field results in the formation 
of a belt-like accretion flow near the equatorial plane. 
The NS spins up as the angular momentum of the accreting gas 
is converted into the angular momentum of the electromagnetic field, 
which then flows into the NS. 
Even if an accretion column forms near one of the magnetic poles, 
the observed luminosity is almost the same on both sides 
with the accretion column and the side without it 
because the radiation energy is transported to both sides through scattering.
Our model suggests that galactic ULXP, 
Swift J0243.6+6124, has a quadrupole magnetic field of 
$2\times10^{13}~{\rm G}$ 
and a dipole magnetic field of less than $4\times10^{12}~{\rm G}$.

\end{abstract}



\section{Introduction} \label{sec:intro}

Ultraluminous X-ray sources (ULXs) are bright, point-like, and non-nuclear X-ray objects whose isotropic X-ray luminosity exceeds the Eddington luminosity for the stellar-mass black holes, $\sim10^{39}~{\rm erg~s^{-1}}$ \citep[see recent reviews by][]{Kaaret2017,Fabrika2021,King2023,Pinto2023}. 
Such a high X-ray luminosity suggests that they host either intermediate-mass  black holes accreting at sub-Eddington rate \citep[see, e.g.,][]{Colbert1999,Makishima2000,Matsumoto2001,Miller2003,Miller2004},  stellar-mass black holes accreting at super-Eddington rate \citep[see, e.g.,][]{King2001,Watarai2001,Poutanen2007,Gladstone2009}, and/or neutron stars (NSs) accreting at super-Eddington rate \citep[see, e.g.,][]{Basko1976,King2001,Bachetti2014,Mushtukov2015b}.
In some ULXs, coherent pulsations with a period of $\sim1-10~{\rm s}$ are detected \citep[e.g.,][]{Bachetti2014,Furst2016,Israel2017a,Israel2017b}.
Such ULXs are called ULX pulsars (ULXPs).
It is widely accepted that the pulsations would be caused by the rotation of the magnetized NS \citep{Mushtukov2018,Inoue2020}.
In order to reconcile the ULX luminosities, super-Eddington accretion onto the magnetized NS is required.

To model the super-Eddington accretion flows, general relativistic radiation magnetohydrodynamics (GR-RMHD) simulations are necessary. So far, many GR-RMHD simulations of super-Eddington accretion flows onto a blackhole are conducted \citep[see, e.g.,][]{Sadowski2014,McKinney2014,Takahashi2016,Utsumi2022,Ricarte2023}. 
These simulations showed that the numerical model can explain the observations of the ULXs, such as the radiative luminosity, kinetic luminosity and radiation spectrum. 
However, only a few GR-RMHD simulations for the magnetized NS case are succeeded \citep[][]{Takahashi2017,Abarca2021,Inoue2023}.

Due to interactions with the gas in the accretion disk, the closed magnetic field lines of the NS open up \citep[see, e.g.,][and references therein]{Parfrey2016}. 
Additionally, within the radius where the magnetic pressure of the NS's magnetic field balances the radiation pressure of the accretion disk (so-called magnetospheric radius), the gas flows onto the NS along its magnetic field lines.
If the dipole magnetic field dominates over the other field components, accretion columns form near the NS's magnetic poles \citep{Takahashi2017,Abarca2021}.
On the other hand, if the quadrupole magnetic field is predominant within the magnetosphere, a belt-like accretion flow appears around the NS's equator.
We refer to such an accretion flow as an accretion belt.
\citet{Long2007} have demonstrated the accretion belt around the star with quadrupole magnetic fields through magnetohydrodynamics (MHD) simulations.
The accretion belt around the NS with a low accretion rate has also been confirmed by \citet{Das2022} through general relativistic magnetohydrodynamic (GR-MHD) simulations.
In both cases, the accretion disks form outside the magnetospheric radius.

Observationally, in the ULXPs, it has been pointed out that the multipole magnetic field, stronger than the dipole magnetic field, exists near the NS surface.
\citet{Israel2017b} suggested that a high luminosity of NGC 5907 ULX $\sim10^{41}~{\rm erg~s^{-1}}$ can be achieved if the multipole magnetic field with a strength of $\sim10^{14}~{\rm G}$ exists \citep[see, also][]{Eksi2015,Brice2021}.
\citet{Kong2022} reported the cyclotron resonance scattering feature (CRSF) with a centroid energy of $E_{\rm cyc}\sim 150~{\rm keV}$ and a line width of $\sigma_{\rm cyc}\sim20-30~{\rm keV}$ in the X-ray spectrum of Swift J0243.6+6124 using data from Insight-HXMT.
The corresponding surface magnetic field strength is $\sim2\times10^{13}~{\rm G}$.
This value is stronger than what is estimated based on the assumption that the NS has a dipole magnetic field \citep[see, e.g.,][]{Tsygankov2018,Doroshenko2020,Inoue2023}.
Furthermore, motivated by the discovery of the CRSF in M51 ULX8 \citep[][]{Brightman2018}, \citet{Middleton2019} also analyzed the X-ray spectra of M51 ULX8.
From the spectral fitting, they obtained two solutions: $(E_{\rm cyc},\sigma_{\rm cyc})\sim (4.5~{\rm keV},0.1~{\rm keV})$ and $(E_{\rm cyc},\sigma_{\rm cyc})\sim (4.5~{\rm keV},1~{\rm keV})$.
Such a CRSF would originate from the resonant scattering by electrons in the dipole magnetic field of $10^{12}~{\rm G}$ or by protons in the multipole magnetic field of $10^{15}~{\rm G}$.

Although the existence of multipolar magnetic fields is frequently discussed, super-Eddington accretion flows onto a NS with multipolar magnetic fields are still not well understood.
In this paper, we investigate the super-Eddington accretion flows around NSs with dipole and/or quadrupole magnetic fields using GR-RMHD simulations.
We demonstrate that the observations, including CRSF, of Swift J0243.6+6124 can be explained by our model if the strength of the dipole magnetic field at the NS's magnetic pole is less than $4\times10^{12}~{\rm G}$ and if that of the quadrupole field is around $2\times10^{13}~{\rm G}$.
This paper is organized as follows: we will present the numerical methods in Section \ref{sec:method} and show the results in Section \ref{sec:result}. 
Section \ref{sec:discussion} is devoted to the discussion of the NS's magnetic field structure in Swift J0243.6+6124.
Finally, we give our conclusion in the final section.

{\section{Method} \label{sec:method}}

We numerically solve the GR-RMHD equations in Schwarzschild polar coordinates $(t,r,\theta,\phi)$ using the numerical code {\tt\string UWABAMI}
\citep{Takahashi2017}.
Based on the moment formalism of the radiation field \citep{Thorne1981}, this code adopts the M1 closure as the closure relation \citep{Levermore1984,Kanno2013,Sadowski2013}.
In this closure, the radiation field is updated by solving the zeroth and first moment of the radiative transfer equation.
The speed of light $c$ and the gravitational constant $G$ are normalized to $1$ unless otherwise specified.
Hereafter, spacetime and space components are represented by Greek and Latin suffixes, respectively.

{\subsection{Basic equations}\label{sec:basic_equations}}

The equations for the time evolution of the GR-RMHD are given by
\citep[see, e.g.,][]{Takahashi2018}
\begin{eqnarray}
    \nabla_\mu\left(\rho u^\mu\right)&=&0,
    \label{eq:mass_cons}
    \\
    \nabla_\mu\left(T^{\mu\nu}\right)&=&G^\nu,
    \label{eq:momentum_cons_gas}
    \\
    \nabla_\mu\left(R^{\mu\nu}\right)&=&-G^\nu,
    \label{eq:momentum_cons_rad}
    \\
    \partial_t\left(\sqrt{-g}B^i\right)&=&-\partial_j\left\{\sqrt{-g}\left(b^i u^j-b^j u^i\right)\right\},
    \label{eq:induction_eq}
\end{eqnarray}
where $\rho$ is the proper mass density, $u^\mu$ is the four-velocity of the gas, $B^i$ is the magnetic field vector in the laboratory frame, $b^\mu$ is the magnetic four-vector in the fluid frame, and $g={\rm det}(g_{\mu\nu})$ is the determinant of the metric.
The energy-momentum tensor of the ideal MHD is $T^{\mu\nu}={T_{\rm MA}}^{\mu\nu}+{T_{\rm EM}}^{\mu\nu}$, where
\begin{eqnarray}
    {T_{\rm MA}}^{\mu\nu}&=&(\rho+e+p_{\rm gas})u^{\mu}u^{\nu}+p_{\rm gas} g^{\mu\nu},
    \label{eq:Tmunu_gas}\\
    {T_{\rm EM}}^{\mu\nu}&=&b^2u^{\mu}u^{\nu}+p_{\rm mag} g^{\mu\nu}-b^\mu b^\nu.\label{eq:_Mmunu}
\end{eqnarray}
Here, $e$ is the internal energy density, $p_{\rm gas}=(\Gamma-1)e$ is the gas pressure ($\Gamma=5/3$), and $p_{\rm mag}=b^2/2$ is the magnetic pressure in the fluid frame.
In the M1 formalism, the energy-momentum tensor of the radiation field is expressed as \citep{Sadowski2013},
\begin{eqnarray}
    R^{\mu\nu}
    =\left(\bar{E}+p_{\rm rad}\right){u_{\rm R}}^{\mu}{u_{\rm R}}^{\nu}
    +p_{\rm rad}g^{\mu\nu}.
    \label{eq:Rmunu}
\end{eqnarray}
Here, $\bar{E}$ is the radiation energy density, $p_{\rm rad}=\bar{E}/3$ is the radiation pressure in the radiation rest frame, and ${u_{\rm R}}^{\mu}$ is the four-velocity of the radiation rest frame.
The interaction between the ideal MHD and radiation field is described by the radiation four-force, 
\begin{eqnarray}
    {G}^\mu =
    &-&{\rho} {\kappa}_{\rm abs}
    \left(R^{\mu\alpha} u_\alpha+4\pi \hat{B} u^\mu\right)\nonumber\\
    &-&{\rho}{\kappa}_{\rm sca}
    \left(R^{\mu\alpha} u_\alpha
    +R^{\alpha\beta} u_\alpha u_\beta u^\mu\right)
    +{G_{\rm comp}}^\mu,
\end{eqnarray}
where $\kappa_{\rm abs}=6.4\times 10^{22}\rho T_{\rm e}^{-3.5}$ ${\rm cm^2~g^{-1}}$ is the opacity for free-free absorption, $\kappa_{\rm sca}=0.4$ ${\rm cm^2~g^{-1}}$ is the isotropic electron scattering opacity, and $T_{\rm e}$ is the electron temperature.
The blackbody intensity is given by $\hat{B}= a{{T}_{\rm e}}^4/4\pi$, where $a$ is the radiation constant.
In this study, we take thermal Comptonization into account \citep{Fragile2018,Utsumi2022}:
\begin{eqnarray}
    {G_{\rm comp}}^\mu
    =-\kappa_{\rm sca}\rho\hat{E}
    \frac{4k(T_{\rm e}-T_{\rm r})}{m_{\rm e}}u^\mu.
    \label{eq:thermal_comp}
\end{eqnarray}
Here, $\hat{E}$ is the radiation energy density in the fluid frame, $T_r=(\hat{E}/a)^{1/4}$ is the radiation temperature, and $m_{\rm e}$ is the electron mass.
We assume $T_{\rm e}=T_{\rm g}$ for simplicity, where $T_{\rm g}$ is the gas temperature.
The gas temperature is calculated from $T_{\rm g}={\mu_{\rm w} m_{\rm p} p_{\rm gas}}/({\rho k})$, where $m_{\rm p}$ is the proton mass, $k$ is the Boltzmann constant, and $\mu_{\rm w}=0.5$ is the mean molecular weight.
We consider the subgrid model to mimic the mean-field dynamo proposed by \citet{Sadowski2015a}.

\begin{deluxetable*}{cccccccccc}[t]
\tablecaption{Parameters for different models\label{tab:table1}}
\tablehead{
\colhead{Parameters}
&\colhead{$f$}
&\colhead{$(N_r,N_\theta)$}
&\colhead{$r_{\rm out}$}
&\colhead{$\rho_0$}
&\colhead{$\dot{M}_{\rm in}$}
&\colhead{$\dot{M}_{\rm out}$}
&\colhead{$L_{\rm rad}$}
&\colhead{$L_{\rm kin}$}
&\colhead{$r_{\rm M}$}\\
\colhead{Unit}
&\colhead{}
&\colhead{}
&\colhead{$[{\rm km}]$}
&\colhead{$[{\rm g~cm^{-3}}]$}
&\colhead{$[\dot{M}_{\rm Edd}]$}
&\colhead{$[\dot{M}_{\rm Edd}]$}
&\colhead{$[L_{\rm Edd}]$}
&\colhead{$[L_{\rm Edd}]$}
&\colhead{$[\rm km]$}
}
\startdata
{\tt\string D\_d001}    &  0  & (608,512) & 840  & 0.01 & 56  & 220   & 8.8 & 0.30 & 28.9\\
{\tt\string DQ\_d001}   & 1/3 & (608,512) & 840  & 0.01 & 59  & 390   & 9.3 & 0.40 & 27.3\\
{\tt\string QD\_d001}   & 2/3 & (608,512) & 840  & 0.01 & 55  & 310   & 21  & 0.47 & 21.2\\
{\tt\string Q\_d001}    &  1  & (608,512) & 840  & 0.01 & 73  & 450   & 14  & 0.53 & 20.4\\
{\tt\string D\_d01}     &  0  & (608,512) & 840  & 0.1  & 530 & 7800  & 86  & 43   & 16.3\\
{\tt\string DQ\_d01}    & 1/3 & (608,512) & 840  & 0.1  & 410 & 10000 & 77  & 34   & 11.2\\
{\tt\string QD\_d01}    & 2/3 & (608,512) & 840  & 0.1  & 290 & 10000 & 70  & 24   & 10.5\\
{\tt\string Q\_d01}     &  1  & (608,512) & 840  & 0.1  & 390 & 11000 & 76  & 25   & 10.3\\
{\tt\string QD\_d01\_a} & 2/3 & (812,512) & 2100 & 0.1  & 370 & 6900  & 72  & 24   & 10.4\\
{\tt\string Q\_d01\_a}  &  1  & (812,512) & 2100 & 0.1  & 500 & 6800  & 81  & 26   & 10.4
\enddata
\tablecomments{
The model names are shown in the first column: the capital letter ``{\tt\string D}'' (``{\tt\string Q}'') means the dipole (quadrupole) magnetic field, and ``{\tt\string dXX}'' denotes the maximum gas density of the initial torus.
The ratio of the magnetic field strength $f=B_{\rm qua}/B_{\rm tot}$, the numerical grid points $(N_r,N_\theta)$, the radius of the outer boundary $r_{\rm out}$, the maximum gas density of the initial torus $\rho_0$, the mass accretion rate $\dot{M}_{\rm in}$, the mass outflow rate $\dot{M}_{\rm out}$, the radiative luminosity $L_{\rm rad}$, the kinetic luminosity $L_{\rm kin}$, and the magnetospheric radius $r_{\rm M}$ are presented.}
\end{deluxetable*}

{\subsection{Numerical models}\label{sec:numrical_model}}

We set the NS's mass and radius to $M_{\rm NS}=1.4M_\odot$ and to $r_{\rm NS}=10~\rm km$, respectively.
The rotation of the NS is ignored because the observed pulse period in ULXPs corresponding to the rotation period of the NS is $1-10~{\rm s}$, which is much longer than the Keplerian timescale, $\sim10^{-2}~{\rm s}$, even within $r\sim100~{\rm km}$.
In the present paper, the axisymmetric system where the magnetic axis coincides with the rotation axis of the accretion disk is assumed.
We fix the total magnetic field strength at the magnetic pole in the upper hemisphere to $B_{\rm tot}=B_{\rm dip}+B_{\rm qua}=4\times10^{10}~{\rm G}$.
Here, $B_{\rm dip}$ and $B_{\rm qua}$ are the dipole and quadrupole magnetic field strength at the NS's magnetic pole, respectively.
Four cases of the magnetic field configuration of the NS parameterized by $f=B_{\rm qua}/B_{\rm tot}=0,1/3,2/3,1$ are investigated.
When $f=0$ ($f=1$), the NS has a pure dipole (quadrupole) magnetic field calculated from $A^{\rm dip}_\phi$ ($A^{\rm qua}_\phi$).
Here, $A^{\rm dip}_\phi$ and $A^{\rm qua}_\phi$ are the vector potential for the dipole and quadrupole magnetic fields, respectively \citep[see, Appendix in][]{Das2022}.
In the cases of $f=1/3,2/3$, we calculate the NS's magnetic field from $A^{\rm dip}_\phi$+$A^{\rm qua}_\phi$ \citep{Long2007}.
The direction of the quadrupole magnetic field is chosen so as to be parallel to that of the dipole one at $(r,\theta)=(r_{\rm NS},0)$.
The computational domain consists of $[r_{\rm NS},r_{\rm out}]$ and $[0,\pi]$, where $r_{\rm out}$ is the radius of the outer boundary.
We run the simulations for $[0,40000t_{\rm g}]$, where $t_{\rm g}$ is the light-crossing time for the gravitational radius of the NS, $r_{\rm g}=M_{\rm NS}=2.1~{\rm km}$.
The size of the radial grid exponentially increases with $r$, while the grids in $\theta$-direction are uniformly distributed \citep{Takahashi2017}.
Table \ref{tab:table1} lists $f$ for each model.
In this table, the numerical grid points $(N_r,N_\theta)$, $r_{\rm out}$, the maximum gas density of the initial torus $\rho_0$, the mass accretion rate $\dot{M}_{\rm in}$, the mass outflow rate $\dot{M}_{\rm out}$, the radiative luminosity $L_{\rm rad}$, the kinetic luminosity $L_{\rm kin}$, and the magnetospheric radius $r_{\rm M}$ are also tabulated.
Here, $\dot{M}_{\rm Edd}=L_{\rm Edd}$ is the Eddington mass accretion rate.
The time-averaged values for $\dot{M}_{\rm in}$, $\dot{M}_{\rm out}$, $L_{\rm rad}$, $L_{\rm kin}$, and $r_{\rm M}$ are presented.
The mass accretion rate and outflow rate are respectively obtained from
\begin{eqnarray}
    \dot{M}_{\rm in}&=&-\int_{r=r_{\rm NS}}\min[\rho u^r,0]\sqrt{-g}d\theta d\phi,\\
    \dot{M}_{\rm out}&=&\int_{r=r_{\rm out}}\max[\rho u^r,0]\sqrt{-g}d\theta d\phi.
    \label{eq:outflow}
\end{eqnarray}
The radiative and kinetic luminosity are respectively calculated from \citep{Sadowski2016}
\begin{eqnarray}
    L_{\rm rad}&=&-\int_{r=r_{\rm out}}\min[R^r_t,0]
    \sqrt{-g}d\theta d\phi,\\
    L_{\rm kin}&=&-\int_{r=r_{\rm out}}\min[\rho u^r \left(u_t+\sqrt{-g_{tt}}\right),0]
    \sqrt{-g}d\theta d\phi.
    \label{eq:kinetic_luminosity}
\end{eqnarray}
In this study, $r_{\rm M}$ is defined as the maximum radius of the region, where the $u_\phi$-weighted $\theta$-average of $(p_{\rm gas}+p_{\rm rad})/p_{\rm mag}$ is smaller than unity \citep{Inoue2023}.
Models {\tt\string QD\_d01\_a} and {\tt\string Q\_d01\_a} have the same initial parameters as models {\tt\string QD\_d01} and {\tt\string Q\_d01}, respectively, except for $r_{\rm out}$.
These models are used to estimate the blackbody radius when the quadrupole magnetic field dominates in the magnetosphere (see Section \ref{sec:discussion} for detail).

We initially put a Fishbone \& Moncrief torus \citep{Fishbone1976} as a source of the accreting gas.
Under the condition of local thermodynamic equilibrium ($T_{\rm g}=T_{\rm r}$), we take $p_{\rm gas}+p_{\rm rad}$ inside the torus instead of $p_{\rm gas}$.
We set the radius of the torus inner edge to $210~{\rm km}$ and the maximum pressure radius to $304.5~{\rm km}$.
In addition to the dipole and quadrupole magnetic fields of the NS, the poloidal-loop magnetic fields whose vector potential is proportional to $\max(\rho/\rho_0-0.2,0)$ are put inside the torus.
The embedded loop magnetic fields are antiparallel to the dipole magnetic field at the torus inner edge \citep{Romanova2011,Takahashi2017,Parfrey2017}.
On the other hand, for $r<304.5~{\rm km}$, the initial loop magnetic fields are antiparallel to quadrupole magnetic fields in the upper hemisphere and parallel to them in the lower hemisphere at the torus surface.
We impose $\max[p_{\rm gas}+p_{\rm rad}]/\max[p_{\rm mag}]=100$ on the loop magnetic field and give a perturbation on $p_{\rm gas}+p_{\rm rad}$ by 10\% to break an equilibrium state.
The NS and torus are surrounded by the relatively hot and low-density corona with a density of $\rho_{\rm col}$ and pressure of $p_{\rm col}$ \citep[see, section 2.2 in][]{Inoue2023}.
The gas velocity of the corona is $u^i=u_{\rm col}^i=0$.
We set the outflowing boundary at $r=r_{\rm out}$ and the reflective boundary at $\theta=0,\pi$.
At $r=r_{\rm NS}$, the gas is swallowed by the NS, but the energy is not swallowed by the NS \citep{Ohsuga2007a,Inoue2023}.

We adopt the simplified version of the method proposed by \citet{Parfrey2017,Parfrey2023} to solve the GR-RMHD equations stably.
The concept of their prescription is to divide the fluid into the contributions from the GR-MHD flows and from the numerical floor.
To do so, we solve $\nabla_\mu\left(\mathcal{F}\rho u^\mu\right)=0$ in addition to equations (\ref{eq:mass_cons})-(\ref{eq:induction_eq}).
Here, $\mathcal{F}$ is evolved as a passive scalar.
We initialize $\mathcal{F}=1$ inside the torus while $\mathcal{F}=0$ outside the torus.
Using $\mathcal{F}$, we adjust the fluid quantities $q=(\rho,p_{\rm gas},u^i)$.
In the region of $\mathcal{F}=0$, we replace $q$ with the quantities of the initial corona $q_{\rm col}=(\rho_{\rm col},p_{\rm col},u_{\rm col}^i)$.
On the other hand, $q$ is not modified in the region of $\mathcal{F}=1$.
When $0<\mathcal{F}<1$, the fluid quantities are linearly interpolated with weight $\mathcal{F}$ between $q_*$ and $q_{\rm col}$ as $q=q_{\rm col}+(q_*-q_{\rm col})\mathcal{F}$.
Here, $q_*=(\rho_*, p_*, u_*^i)$ are the gas density ($\rho_*$), pressure ($p_*$), and velocity ($u_*^i$) calculated from the conservative variables.
We take $\kappa_{\rm sca}=0$ and $\kappa_{\rm abs}=0$ in the regions where $\mathcal{F}<0.9$ and $\sigma=b^2/\rho>10$ for the numerical stability.

{\section{Result} \label{sec:result}}

\subsection{Accretion Structure\label{sec:accretion_flows}}

\begin{figure*}[tb]
\centering
\includegraphics[width=\linewidth]{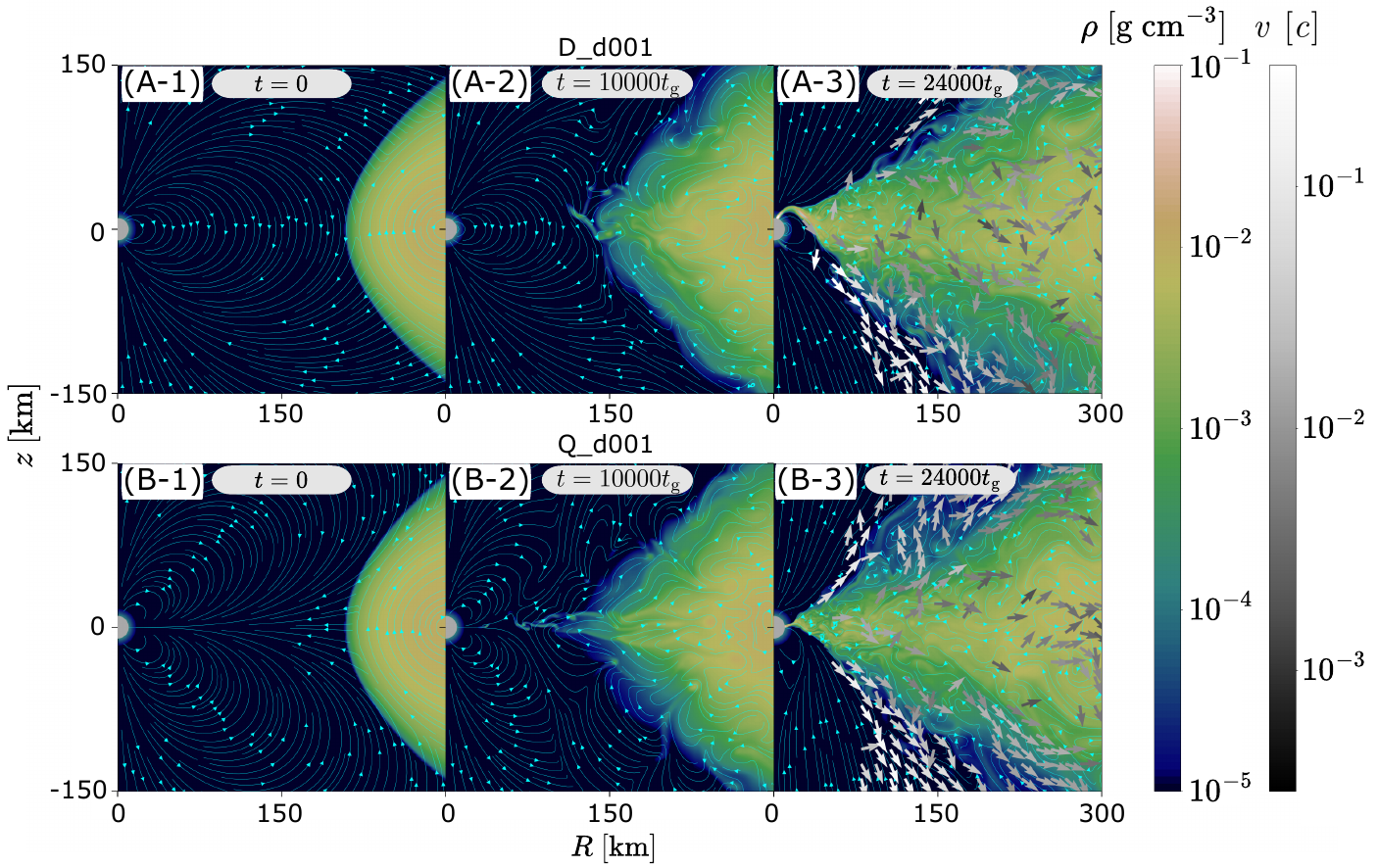}
\caption{
Gas density distribution at $t=0,~10000t_{\rm g},~24000t_{\rm g}$ for models {\tt\string D\_d001} and {\tt\string Q\_d001}.
Stream plots are magnetic field vectors.
Vectors in panels (A-3) and (B-3) are the gas velocity vectors in the region where $4\pi r^2 \rho u^r>\dot{M}_{\rm Edd}$.
Here, the color of the vectors indicates the magnitude of the velocity in the poloidal direction.
The grey region of $r<10~{\rm km}$ is the NS.
\label{fig:figure1}}
\end{figure*}

In Figure \ref{fig:figure1}, we describe the time evolution of the gas density distribution and the magnetic field.
Here, $(R,z)=(r\sin\theta,r\cos\theta)$.
We present the results of models {\tt\string D\_d001} (upper panels) and {\tt\string Q\_d001} (lower panels).
The NS is depicted by the grey region of $r<10~{\rm km}$, and its center is located at the origin $(R,z)=(0,0)$.
Stream plots represent the magnetic field vectors.
Vectors in panels (A-3) and (B-3) are the gas velocity vectors in the poloidal direction, which are plotted only in the region where $4\pi r^2 \rho u^r>\dot{M}_{\rm Edd}$.
The color of the vectors indicates the magnitude of the velocity in the poloidal direction.
As mentioned in Section \ref{sec:method}, the dipole magnetic field is antiparallel to the loop magnetic field at the inner edge of the torus (see, panel (A-1)).
On the other hand, the quadrupole magnetic field is antiparallel (parallel) to the loop magnetic field at the torus surface in the upper (lower) hemisphere (see, panel (B-1)).

After the onset of the simulations, the system starts to deviate from the equilibrium state.
The gas goes inward due to the Maxwell stress induced by the magnetorotational instability \citep[MRI;][]{Balbus1991} (panels (A-2) and (B-2)).
The embedded loop magnetic fields reconnect with the initially closed magnetic fields of the NS, causing the NS's magnetic field lines to open up.
The accretion disk is formed near the equatorial plane, $z=0$ (see panels (A-3) and (B-3)).

As can be seen from the velocity vectors, outflows emanate from the accretion disk.
Such outflows are driven by the radiation force and centrifugal force and appear in all models.
Although not shown in this figure, the effective optical depth of the outflows measured from the outer boundary in the $r$-direction exceeds unity \citep[][]{Ogawa2017,Inoue2023}.
This implies that the outflows from the super-Eddington accretion disk produce thermal emission, regardless of whether the NS’s magnetic field is a dipole or quadrupole.
Furthermore, when estimated using the same method as \citet{Inoue2023}, the blackbody radii are consistent with the relation of $r_{\rm  bb}=3.2(\dot{M}_{\rm in}/\dot{M}_{\rm Edd})^{0.71}$ in \citet[][]{Inoue2023}, even if the quadrupole magnetic field dominates inside the magnetosphere ($51~{\rm km}$ in model {\tt\string Q\_d001}, $170~{\rm km}$ in model {\tt\string QD\_d01\_a}, and $160~{\rm km}$ in model {\tt\string Q\_d01\_a}).
The magnitude of the poloidal velocity in the regions around $\theta\sim45^\circ$ and $135^\circ$ is larger than that in the range of $45^\circ\lesssim\theta\lesssim 135^\circ$ and exceeds $0.1c$. 
Since the gas density in these regions is very low, the radiation from the accretion flows effectively accelerates the outflowing gas \citep[][]{Ohsuga2007c}.

The accretion flows along the dipole or quadrupole magnetic field lines exist near the NS.
The details of these structures will be explained later.
In models of $\rho_0=0.01 {\rm g~cm^{-3}}$, the radiative shock arises above the NS surface (see Appendix \ref{sec:shock} for details).
On the other hand, such a structure cannot be seen in models of $\rho_0=0.1 {\rm g~cm^{-3}}$.
The past numerical simulations carefully examined the radiative shock structure \citep{Kawashima2016,Kawashima2020,Zhang2022,Zhang2023,Abolmasov2023}, but the detailed structure is out of the scope of this study.

We should stress here that the super-Eddington accretion is feasible since a large amount of the radiation energy escapes from the side wall of the accretion column (see, Appendix \ref{sec:radiation_from_ac_ab}), which has been shown by \citet{Kawashima2016}.
It reduces the radiation energy density inside the column ($\hat{E}=u_\mu u_\nu R^{\mu\nu}$), leading to a reduction of the outward radiation flux in the fluid frame, $\hat{F}_{\rm rad}\sim(c/\tau)\hat{E}$.
The resulting order of $\kappa_{\rm es}\hat{F}_{\rm rad}/g$ ($g=M_{\rm NS}/r^2$) inside the column is unity \citep[see, also][]{Kawashima2016}, which reduces the gas infall velocity but is not sufficient to prevent the gas accretion.
Thus, super-Eddington accretion onto the magnetized NS through the accretion columns is feasible.

\begin{figure*}[tb]
\centering
\includegraphics[width=\linewidth]{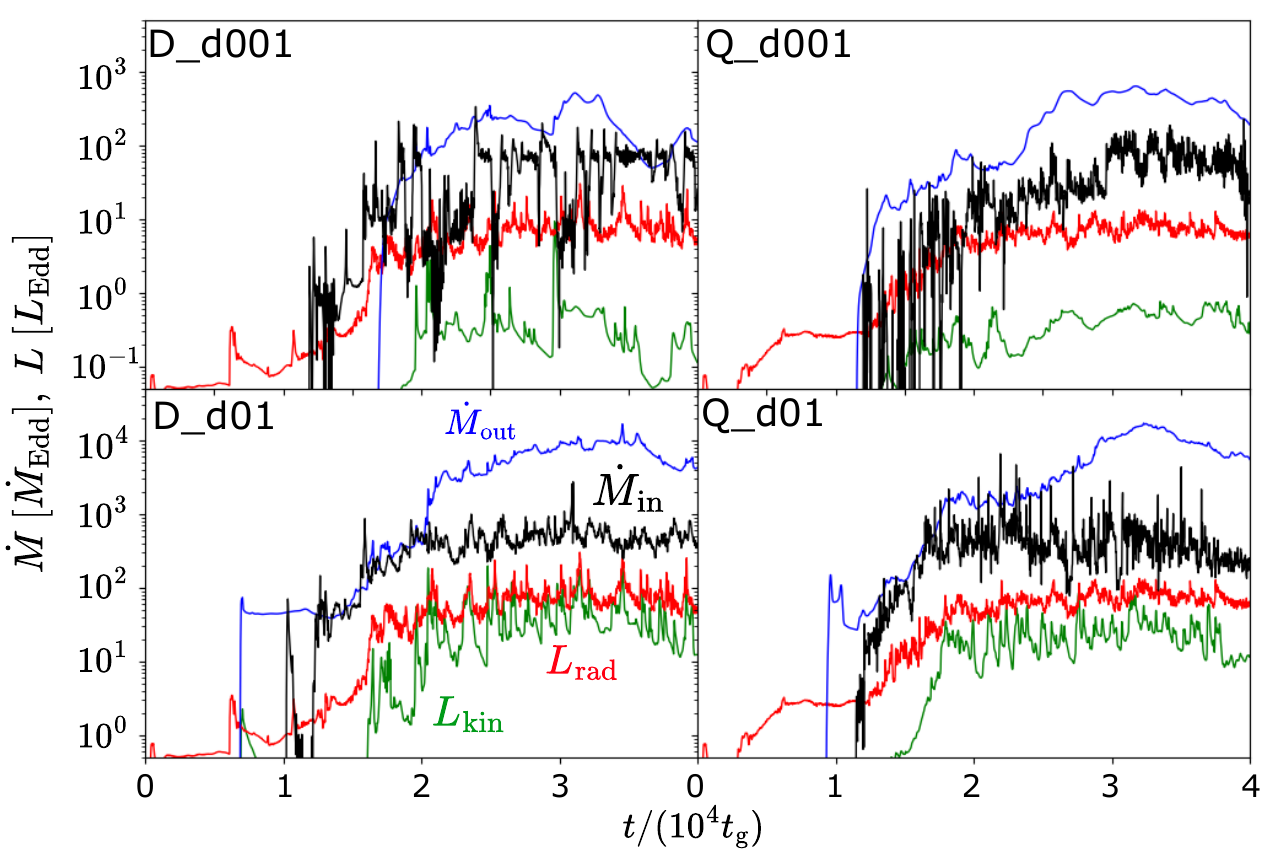}
\caption{Time evolution of the mass accretion rate (black), outflow rate (blue), radiative luminosity (red), and kinetic luminosity (green).
\label{fig:figure2}}
\end{figure*}

We plot $\dot{M}_{\rm in}$, $\dot{M}_{\rm out}$, $L_{\rm rad}$, and $L_{\rm kin}$ as a function of time in Figure \ref{fig:figure2}.
Here, the results of models {\tt\string D\_d001}, {\tt\string Q\_d001}, {\tt\string D\_d01}, and {\tt\string Q\_d01} are presented.
It can be seen that $\dot{M}_{\rm in}$, $\dot{M}_{\rm out}$, $L_{\rm rad}$, and $L_{\rm kin}$ gradually increase after the simulations start, and these quantities are in the quasi-steady state for $[30000t_{\rm g},40000t_{\rm g}]$.
Such time evolution profiles are true for $\dot{M}_{\rm in}$, $\dot{M}_{\rm out}$, $L_{\rm rad}$, and $L_{\rm kin}$ of the other models.
Therefore, we hereafter show the time-averaged results in $[30000t_{\rm g},40000t_{\rm g}]$.
In this time interval, the inflow-outflow structures in all models is in quasi-steady state within $r\sim 100~{\rm km}$ (i.e., net flow rate $\dot{M}_{\rm in}-\dot{M}_{\rm out}$ is almost constant within $r\sim100~{\rm km}$).

We find that $\dot{M}_{\rm in}$, $\dot{M}_{\rm out}$, $L_{\rm rad}$, and $L_{\rm kin}$ in models of $\rho_0=0.1 {\rm g~cm^{-3}}$ are larger than those in models of $\rho_0=0.01 {\rm g~cm^{-3}}$ (see, also Table \ref{tab:table1}).
On the other hand, we cannot find any clear dependence of $\dot{M}_{\rm in}$, $\dot{M}_{\rm out}$, $L_{\rm rad}$, and $L_{\rm kin}$ on $f$.
We can also see that $\dot{M}_{\rm out}$ in models {\tt\string QD\_d01} and {\tt\string Q\_d01} is slightly greater than in models {\tt\string QD\_d01\_a} and {\tt\string Q\_d01\_a}.
This difference arises from the small $r_{\rm out}$ in models {\tt\string QD\_d01} and {\tt\string Q\_d01}.
In these models, the radial gas velocity for $45^\circ\lesssim \theta\lesssim135^\circ$ is smaller than the escape velocity, even at $r_{\rm out}=840~{\rm km}$.
Although a fraction of such gases do not reach $r=2100~{\rm km}$, all gas at $r_{\rm out}$ is included in the integration of equation (\ref{eq:outflow}).
It leads to a larger $\dot{M}_{\rm out}$ in models {\tt\string QD\_d01} and {\tt\string Q\_d01} compared to that in models {\tt\string QD\_d01\_a} and {\tt\string Q\_d01\_a}.

\begin{figure*}[tb]
\centering
\includegraphics[width=0.6\linewidth]{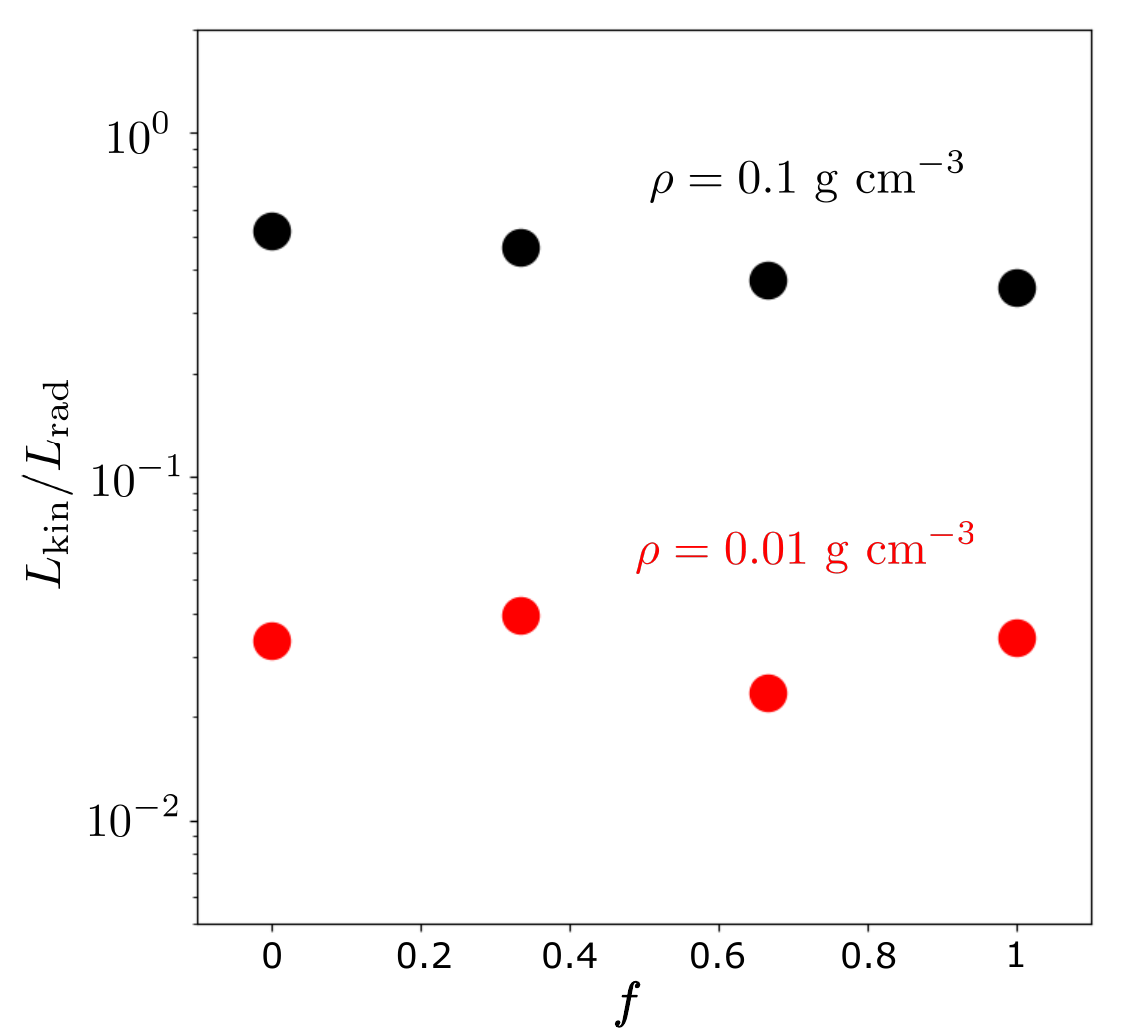}
\caption{The ratio of the kinetic luminosity to the radiative luminosity as a function of $f$.
\label{fig:Lkin_Lrad}}
\end{figure*}

Figure \ref{fig:Lkin_Lrad} shows $L_{\rm kin}/L_{\rm rad}$ as a function of $f$.
We can see that $L_{\rm kin}/L_{\rm rad}$ in models of $\rho_0=0.1 {\rm g~cm^{-3}}$ is about ten times greater than that in models of $\rho_0=0.01 {\rm g~cm^{-3}}$.
This trend is also reported in past numerical simulations \citep{Ohsuga2007a,Inoue2023}.
We cannot find the dependence of $L_{\rm kin}/L_{\rm rad}$ on $f$.

\begin{figure*}[tb]
\centering
\includegraphics[width=\linewidth]{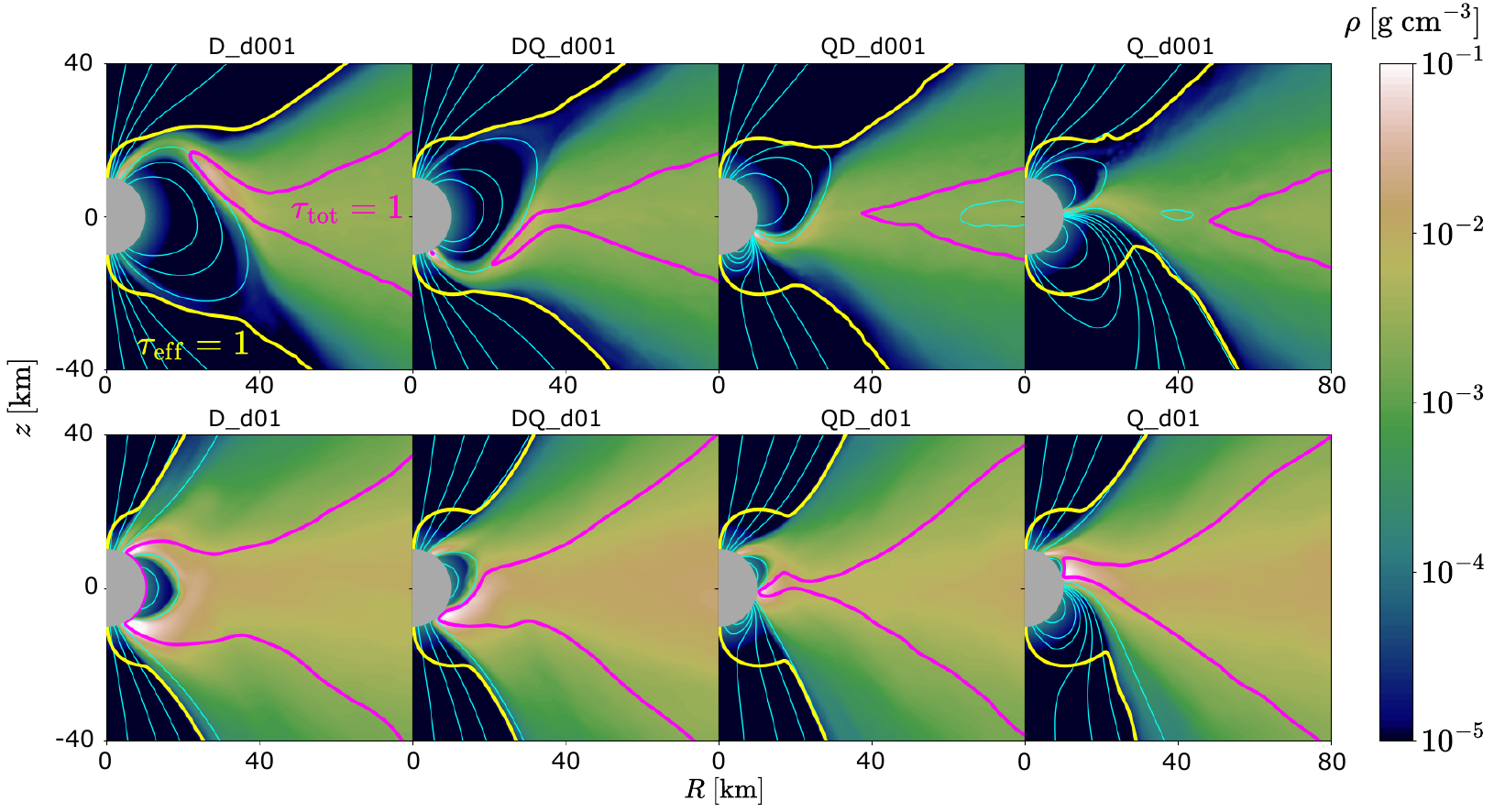}
\caption{Time-averaged gas density map with magnetic field lines (cyan solid lines).
Yellow lines and magenta lines represent the photosphere for the total optical depth and effective absorption, respectively.
\label{fig:figure3}}
\end{figure*}

Next, we explain that the accretion flows along the dipole (quadrupole) magnetic field lines are formed near the NS surface when the dipole (quadrupole) magnetic fields are dominant at $r_{\rm M}$.
Figure \ref{fig:figure3} illustrates the time-averaged gas density (color) and magnetic field lines (cyan solid lines).
In models {\tt\string D\_d001}, {\tt\string DQ\_d001}, {\tt\string QD\_d001}, {\tt\string D\_d01}, and {\tt\string DQ\_d01}, the accretion flows roughly follow the dipole magnetic field lines (we refer to such accretion flows as dipolar accretion flows).
The accretion disk is truncated by the NS's magnetic field and accretion columns are formed near the magnetic poles of the NS,
at $(R,z)\sim(4{\rm km},11{\rm km})$ for model {\tt\string D\_d001}, at $(R,z)\sim(6{\rm km},-11{\rm km})$ for model {\tt\string DQ\_d001}, at $(R,z)\sim(11{\rm km},-5{\rm km})$ for model {\tt\string QD\_d001}, at $(R,z)\sim(6{\rm km},\pm10{\rm km})$ for model {\tt\string D\_d01}, and at $(R,z)\sim(7{\rm km},10{\rm km})$ and $\sim(9{\rm km},-8{\rm km})$ for model {\tt\string DQ\_d01}.
When we define the truncation radius as the radius at which $\sigma=1$ on the equatorial plane, the truncation radius coincides with the magnetospheric radius $r_{\rm M}$ within a factor of two.

In models {\tt\string Q\_d001}, {\tt\string QD\_d01}, and {\tt\string Q\_d01}, the accretion flows along the quadrupole magnetic field lines can be seen.
We name such accretion flows quadrupolar accretion flows.
In these models, $p_{\rm dip}(r_{\rm M})<p_{\rm qua}(r_{\rm M})$ holds ($p_{\rm dip}/p_{\rm qua}\sim0.26$ for model {\tt\string QD\_d001}).
Here, $p_{\rm dip}=B_{\rm dip}^2/(8\pi)(r_{\rm NS}/r)^6$ and $p_{\rm qua}=B_{\rm qua}^2/(8\pi)(r_{\rm NS}/r)^8$ are the magnetic pressure originating from the dipole and quadrupole magnetic field, respectively.
The high-density region existing around the equatorial plane within $r_{\rm M}$ is an accretion belt \citep[][]{Long2007,Das2022}.
As illustrated in Figure \ref{fig:figure3}, such an accretion belt appears at $(R,z)\sim(12{\rm km},1{\rm km})$ for model {\tt\string Q\_d001}, at $(R,z)\sim(10{\rm km},0{\rm km})$ for model {\tt\string QD\_d01}, and at $(R,z)\sim(10{\rm km},2{\rm km})$ for model {\tt\string Q\_d01}.
The angular momentum of the accreting gas within $r_{\rm M}$ is about one-tenth of the Keplerian angular momentum.
The accretion belt has lost angular momentum due to the quadrupole magnetic field and exhibits different properties compared to the accretion disk outside the magnetospheric radius (see, Section \ref{sec:angular_momentum_transfer}).
In the quadrupolar accretion flow models of relatively high-mass accretion rate ({\tt\string QD\_d01} and {\tt\string Q\_d01}), the accretion column is also formed at $(R,z)\sim(7{\rm km},10{\rm km})$.
The accreting matters do not reach the NS surface in the lower hemisphere in models {\tt\string Q\_d001}, {\tt\string QD\_d01}, and {\tt\string Q\_d01}.
The reason for this is that the open magnetic field lines reaching the outer boundary in the lower hemisphere prevent the gas from falling onto the lower hemisphere of the NS \citep[see also][]{Das2022}.

For a fixed $\dot{M}_{\rm in}$, the magnetospheric radius $r_{\rm M}$ tends to be small for the models of a large $f$ (see, Table \ref{tab:table1}).
The reason for this is that $p_{\rm qua}(r)$ decreases with distance from the NS more rapidly than $p_{\rm dip}(r)$.
The magnetospheric radius in models of $\rho_0=0.1 {\rm g~cm^{-3}}$ is smaller than that in models of $\rho_0=0.01 {\rm g~cm^{-3}}$.
This is because the larger the mass accretion rate, the larger the radiation pressure, which leads to a smaller $r_{\rm M}$ for a fixed magnetic pressure.

When the gas in the disk reaches at $R=r_{\rm M}$, it moves along the last closed field line toward the lower gravitational potential. 
As a result, a single accretion stream is formed in models of $\rho_0=0.01 ~{\rm g~cm^{-3}}$ (upper panels). 
\footnote{
In the case of model {D\_d001}, 
despite the NS having only a dipole magnetic field, gas falls onto only one of the poles. The reason is as follows.
As the gas from the torus reaches the magnetosphere, it accretes toward one of the poles, and which pole the gas falls onto is determined by the embedded perturbation in the torus. 
Then, the ram pressure of the gas distorts the last closed field line and tilts it with respect to the equatorial plane (see the last closed magnetic field line depicted in the panel of {\tt\string D\_d001}). 
As the magnetic field lines tilt, the subsequent gas tends to flow in the direction in which the preceding gas accreted. Thus, an accretion column forms on the side where the gas first accretes.
}
On the other hand, in models of $\rho_0=0.1 ~{\rm g~cm^{-3}}$, $R=r_{\rm M}$ is close to NS and the disk is geometrically thick, so that two flows moving toward the direction of low gravitational potential appear along the last closed field line (dual accretion stream) as shown in lower panels.

The photospheres for scattering and effective absorption, which are integrated from the rotational axis, are shown by yellow and magenta lines, respectively.
Here, we define total ($\tau_{\rm tot}$) and effective optical depth ($\tau_{\rm eff}$) as follows:
\begin{eqnarray}
    \tau_{\rm tot}&=&\int\rho(\kappa_{\rm abs}+\kappa_{\rm es}) \sqrt{g_{\theta\theta}}d\theta,\\
    \tau_{\rm eff}&=&\int\rho\sqrt{(\kappa_{\rm abs}+\kappa_{\rm es})\kappa_{\rm abs}} \sqrt{g_{\theta\theta}}d\theta.    
\end{eqnarray}
In models of $\rho_0=0.01 {\rm g~cm^{-3}}$, the larger $f$, the larger $r_{\rm eff}^{\rm min}$.
Here, $r_{\rm eff}^{\rm min}$ is the minimum $r$ for the region of $\tau_{\rm eff}>1$ (the region enclosed by the magenta line).
For instance, $r_{\rm eff}^{\rm min}\sim20 {\rm km}$ for model {\tt\string D\_d001}, while $r_{\rm eff}^{\rm min}\sim40 {\rm km}$ for model {\tt\string Q\_d001}.
This tendency arises from the fact that the disk gas density decreases as $f$ increases.
In model {\tt\string D\_d001}, the relatively high-density disk exists since the accreting gas accumulates around $r=r_{\rm M}$ near the equatorial plane due to the magnetic pressure originating from the dipole magnetic field.
It leads to the large $\kappa_{\rm abs}$, resulting in the small $r_{\rm eff}^{\rm min}$.
Actually, although the mass accretion rate at $r=40 {\rm km}$ is $\sim100\dot{M}_{\rm Edd}$ in both models, $(v^{(r)},\Sigma)\sim(10^{-3}c,10^4~{\rm g~cm^{-2}})$ for model {\tt\string D\_d001} and $(v^{(r)},\Sigma)\sim(10^{-2} c,10^3~{\rm g~cm^{-2}})$ for model {\tt\string Q\_d001}.
Here, $v^{(r)}=u^{(r)}/u^{(t)}$ is the radial gas velocity, where the parentheses denote the quantities in the static observer frame, and $\Sigma=\int_0^\pi \rho\sqrt{g_{\theta\theta}}d\theta$ is the surface gas density.
It is also obvious that $r_{\rm eff}^{\rm min}$ for models {\tt\string DQ\_d001} and {\tt\string QD\_d001} are larger (smaller) than that for model {\tt\string D\_d001} ({\tt\string Q\_d001}).
In models of $\rho_0=0.1 ~{\rm g~cm^{-3}}$, the gas density is high enough for $r_{\rm eff}^{\rm min}=r_{\rm NS}$.

The opening angle of the photosphere for scattering in models of $\rho_0=0.1 ~{\rm g~cm^{-3}}$ is smaller than in models of $\rho_0=0.01 ~{\rm g~cm^{-3}}$.
This is because the gas density of the outflows gets larger as the mass accretion rate increases.
In all models, we expect that the observed radiation spectra are affected by Comptonization since the total optical depth $\tau_{\rm tot}$ at the effective photosphere highly exceeds 100.
The gas temperature weighted by $\Delta\tau_{\rm es}=\rho\kappa_{\rm es}\sqrt{g_{\theta\theta}}\Delta\theta$ is averaged over $\theta$ within the region where $\tau_{\rm tot}>1$ and $\tau_{\rm eff}<1$, resulting in $\sim10^7~{\rm K}$.
It leads to the Compton y-parameter $y=(4kT_{\rm e}/mc^2)\tau_{\rm tot}^2$ larger than unity (see, discussion for detail).

\begin{figure*}[tb]
\centering
\includegraphics[width=0.6\linewidth]{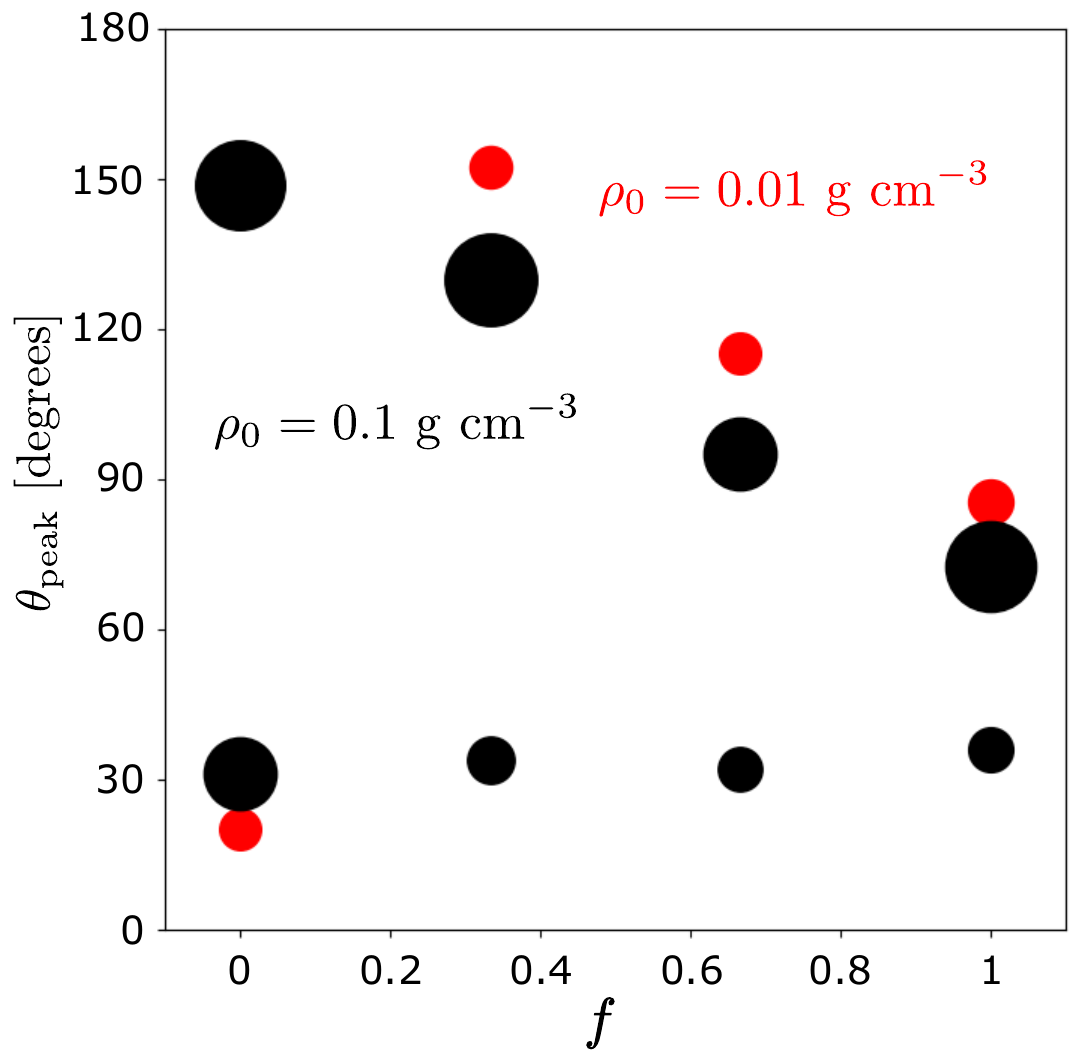}
\caption{
{The position angle ($\theta_{\rm peak}$) of the accretion column and belt as a function of $f$.
Here, we define $\theta_{\rm peak}$ as the polar angle at which $-2\pi r^2\rho u^r$ is maximum in each region of $0^\circ<\theta<60^\circ$ and $60^\circ<\theta<180^\circ$.
The marker size is proportional to the mass accretion rate of the column or belt $\dot{M}_{\rm in}^{\rm upper(lower)}$ (see equation \ref{eq:upper_lower_Mdot}).
The magnetic pole on the upper (lower) hemisphere corresponds to $0^\circ$ ($180^\circ$), and $90^\circ$ represents the equatorial plane.}
\label{fig:figure4}}
\end{figure*}

In Figure \ref{fig:figure4}, the polar angle of the accretion flows at the NS surface $\theta_{\rm peak}$ as a function of $f$ is shown.
Here, $\theta_{\rm peak}$ is defined as the polar angle where $-2\pi r^2 \rho u^r\sin\theta$ is maximum in each region of $0^\circ<\theta<60^\circ$ and $60^\circ<\theta<180^\circ$.
The accretion flow formed in the region of $\mathcal{F}<0.9$ or $\sigma>10$ is ignored where we artificially reduce the emission and absorption processed for numerical stability.
The marker size is proportional to the mass accretion rate at the NS surface integrated within $\theta_{\rm peak}-15^\circ<\theta<\theta_{\rm peak}+15^\circ$,
\begin{eqnarray}
    \dot{M}_{\rm in}^{\rm upper(lower)}=-2\pi\int_{\theta_{\rm peak}-15^\circ}^{\theta_{\rm peak}+15^\circ}\min[\rho u^r,0]\sqrt{-g}d\theta,
    \label{eq:upper_lower_Mdot}
\end{eqnarray}
where the mass accretion rate of the accretion flow closer to $\theta=0^\circ$ $(180^\circ)$ is denoted by ``upper'' (``lower'').
Table \ref{tab:table2} lists $\dot{M}_{\rm in}^{\rm upper}$ and $\dot{M}_{\rm in}^{\rm lower}$.
In this table, $L_{\rm rad}^{\rm upper}$ and $L_{\rm rad}^{\rm lower}$ are also presented.
Here, $L_{\rm rad}^{\rm upper}$ $(L_{\rm rad}^{\rm lower})$ is the radiative luminosity obtained by integrating over the spherical surface of the upper (lower) hemispheres at $r=r_{\rm out}$.
The single accretion stream is formed when $\rho_0=0.01 ~{\rm g~cm^{-3}}$.
Therefore, there is one point (red point) for a given $f$, and $\dot{M}_{\rm in}^{\rm lower}$ cannot be defined.
On the other hand, in the case of $\rho_0=0.1 ~{\rm g~cm^{-3}}$, dual accretion streams appear, so two points (black points) are plotted for a given $f$.
The polar angle of the high-density region near the NS surface illustrated in Figure \ref{fig:figure3} is consistent with $\theta_{\rm peak}$.
In models of $\rho_0=0.01 ~{\rm g~cm^{-3}}$, $\theta_{\rm peak}$ approaches $90^\circ$ as $f$ increases.
To give a specific example, $\theta_{\rm peak}\sim30^\circ$ for model {\tt\string D\_d001}, and $\theta_{\rm peak}\sim90^\circ$ for model {\tt\string Q\_d001}.
When $\rho_0=0.1 {\rm g~cm^{-3}}$,  $\theta_{\rm peak}$ closer to $\theta=180^\circ$ decreases as $f$ increases.
Actually, $\theta_{\rm peak}$ closer to $\theta=180^\circ$ is $\sim150^\circ$ for model {\tt\string D\_d01} and $\sim70^\circ$ for model {\tt\string Q\_d01}.
In these models, $\theta_{\rm peak}$ closer to $\theta=0^\circ$ is $\theta_{\rm peak}\sim30^\circ$ and does not so depend on $f$.

When comparing models of the same $f$ with each other, $\theta_{\rm peak}$ for the model of $\rho_0=0.01 ~{\rm g~cm^{-3}}$ is either greater than the larger $\theta_{\rm peak}$ or less than the smaller $\theta_{\rm peak}$ for the model of $\rho_0=0.1 ~{\rm g~cm^{-3}}$. 
This arises from the fact that the magnetospheric radius in the low-mass accretion rate models is larger than that in the high-mass accretion rate models.
As the magnetospheric radius increases, the last closed magnetic field line connects the polar angle closer to the NS's magnetic pole.
This results in gas accreting at an angle nearer to the magnetic poles.
The result that $\theta_{\rm peak}$ depends on both $f$ and $\rho_0$ implies that the angular distribution of the radiation flux near the NS is also dependent on both $f$ and $\rho_0$.
This point will be explained in Section \ref{sec:radiation_flux}.

Since the gas density distribution in model {\tt\string D\_d01} is approximately symmetric with respect to the equator, $\dot{M}_{\rm in}^{\rm upper}$ is comparable to $\dot{M}_{\rm in}^{\rm lower}$.
On the other hand, $\dot{M}_{\rm in}^{\rm lower}\sim(3-4)\dot{M}_{\rm in}^{\rm upper}$ in models {\tt\string DQ\_d01}, {\tt\string QD\_d01}, and {\tt\string Q\_d01}.
This is because the NS's magnetic field deviates from the dipole field due to the quadrupole component, making it easier for gas to accrete mainly in the region of $60^\circ<\theta<180^\circ$.
Despite the asymmetric accretion flows with respect to the equatorial plane in models {\tt\string D\_d001}, {\tt\string DQ\_d001}, {\tt\string QD\_d001}, {\tt\string DQ\_d01}, {\tt\string QD\_d01}, and {\tt\string Q\_d01}, $L_{\rm rad}^{\rm upper}\sim L_{\rm rad}^{\rm lower}$ holds in all models, and the difference between $L_{\rm rad}^{\rm upper}$ and $L_{\rm rad}^{\rm lower}$ is at most twice.
The reason for this will be presented in Section \ref{sec:radiation_flux}.

\begin{deluxetable}{ccccc}[t]
\tablecaption{The accretion rate and radiative luminosity\label{tab:table2}}
\tablehead{
\colhead{}
&\colhead{$\dot{M}_{\rm in}^{\rm upper}$}
&\colhead{$\dot{M}_{\rm in}^{\rm lower}$}
&\colhead{$L_{\rm rad}^{\rm upper}$}
&\colhead{$L_{\rm rad}^{\rm lower}$}\\
\colhead{Model}
&\colhead{$[\dot{M}_{\rm Edd}]$}
&\colhead{$[\dot{M}_{\rm Edd}]$}
&\colhead{$[L_{\rm Edd}]$}
&\colhead{$[L_{\rm Edd}]$}
}
\startdata
{\tt\string D\_d001}  & 20  & --- & 2.4 & 5.9 \\
{\tt\string DQ\_d001} & 58  & --- & 5.3 & 3.6 \\
{\tt\string QD\_d001} & 55  & --- & 6.7 & 13  \\
{\tt\string Q\_d001}  & 73  & --- & 7.5 & 6.3 \\
{\tt\string D\_d01}   & 220 & 310 & 48 & 34   \\
{\tt\string DQ\_d01}  & 84  & 330 & 43 & 29   \\
{\tt\string QD\_d01}  & 68  & 220 & 36 & 29   \\
{\tt\string Q\_d01}   & 71  & 310 & 33 & 37
\enddata
\end{deluxetable}

\begin{figure*}[tb]
\centering
\includegraphics[width=\linewidth]{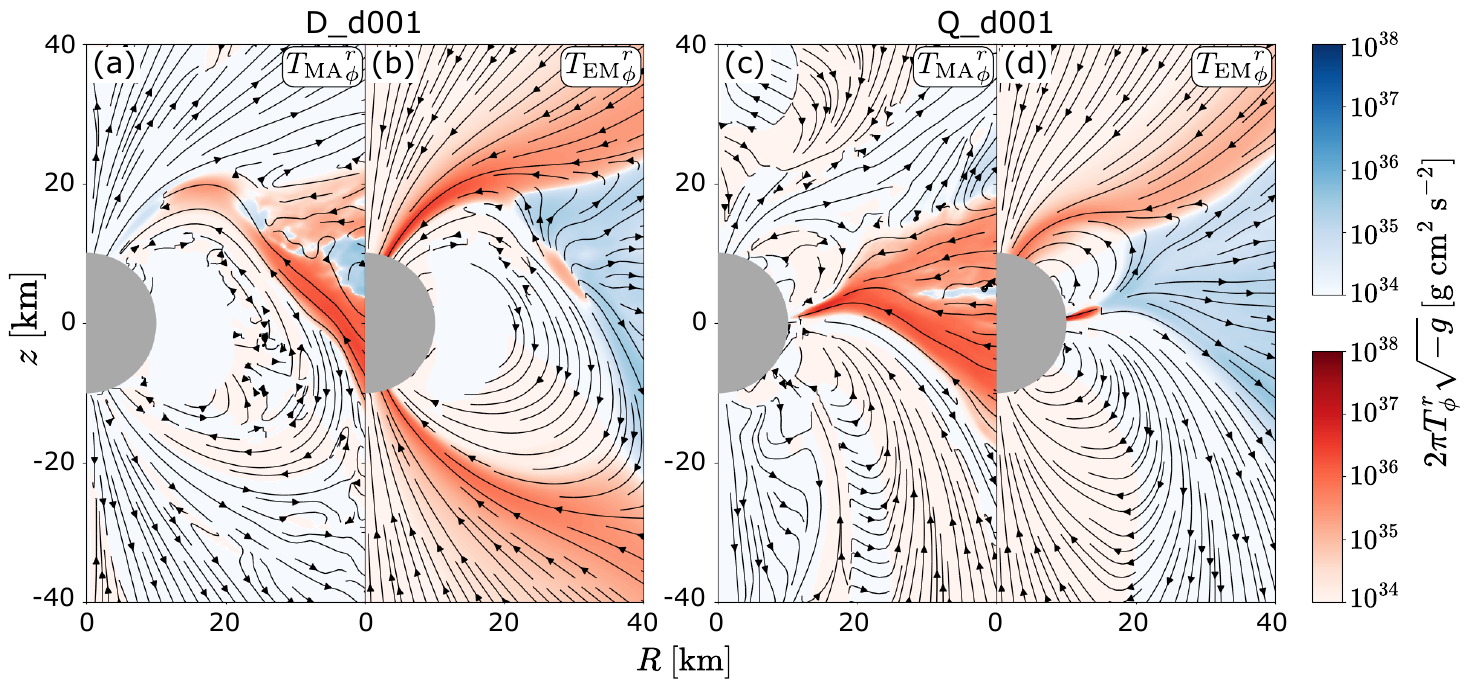}
\caption{Color contour plots of the angular momentum flux.
The red (blue) region represents the negative (positive) angular momentum flux.
Black vectors depict the poloidal components of the angular momentum flux.
\label{fig:figure5}}
\end{figure*}

\subsection{Angular momentum transfer\label{sec:angular_momentum_transfer}}

Next, we demonstrate that the angular momentum of the accreting gas is transported to the NS via the dipole or quadrupole fields.
Figure \ref{fig:figure5} displays the distribution of the angular momentum flux for models {\tt\string D\_d001} ($r_{\rm M}\sim29~{\rm km}$) and {\tt\string Q\_d001} ($r_{\rm M}\sim20~{\rm km}$).
The red region represents the inward (negative) angular momentum flux, while the outward (positive) angular momentum flux is illustrated in the blue region.
Black vectors in panels (a,c) and (b,d) are the vectors of $({T_{\rm MA}}_\phi^{(r)},{T_{\rm MA}}_\phi^{(\theta)})$ and $({T_{\rm EM}}_\phi^{(r)},{T_{\rm EM}}_\phi^{(\theta)})$, respectively. 
In model {\tt\string D\_d001}, ${T_{\rm MA}}_\phi^r$ is negative inside the disk region, which indicates that the angular momentum is transported inward due to the gas accretion (see panel (a)).
It can also be seen that $|{T_{\rm MA}}_\phi^r|$ for $R\lesssim r_{\rm M}$ is smaller than $|{T_{\rm MA}}_\phi^r|$ for $R\gtrsim r_{\rm M}$.
The reason for the decrease in $|{T_{\rm MA}}_\phi^r|$ is that the inward ${T_{\rm MA}}_\phi^r$ is converted into the inward ${T_{\rm EM}}_\phi^r$ at $R\sim r_{\rm M}$ via the interaction between the dipole magnetic field and the accreting matter.
Actually, the inward ${T_{\rm EM}}_\phi^r$ appears for $R\lesssim r_{\rm M}$, and the black vectors point from $(R,z)\sim(r_{\rm M},0)$ to $(R,z)\sim(5~{\rm km},\pm8~{\rm km})$ (see panel (b)).
This inward ${T_{\rm EM}}_\phi^r$ leads the NS to spin up.
Whereas the mass accretion rate at the NS surface in the lower hemisphere is small, the inward angular momentum flux is significant near both magnetic poles.
In fact, $2\pi{T_{\rm EM}}_\phi^{r}\sqrt{-g}\sim 5\times10^{36}~{\rm g~cm^{2}~s^{-2}}$ at $(R,z)\sim(5~{\rm km},8~{\rm km})$, and $2\pi{T_{\rm EM}}_\phi^{r}\sqrt{-g}\sim 2\times10^{36}{\rm g~cm^{2}~s^{-2}}$ at $(R,z)\sim(5~{\rm km},-8~{\rm km})$.
Since the upper and lower hemispheres of the NS connect each other by magnetic field lines, the angular momentum transferred by the gas accretion around $R\sim r_{\rm M}$ is conveyed to both hemispheres through the magnetic field.

In model {\tt\string Q\_d001} as well, the angular momentum of the accreting gas is transferred to the NS via the quadrupole magnetic field.
We find that ${T_{\rm MA}}_\phi^r$ is negative in the disk region, but its absolute value decreases at around $(R,z)=(10~{\rm km},0)$, where it is replaced by a significant inward ${T_{\rm EM}}_\phi^r$ (see panel (d)).
In addition, the angular momentum is also transferred to NS around $(R,z)\sim(5~{\rm km},8~{\rm km})$, where the mass accretion rate is low. 
This structure is similar to the lower hemisphere in model {\tt\string D\_d001}.
The angular momentum transport can happen even outside accretion columns and accretion belts.
In the disk region for both models {\tt\string D\_d001} and {\tt\string Q\_d001}, ${T_{\rm EM}}_\phi^r$ is positive.
In this region, the angular momentum is transported outward due to the MRI turbulence, which enables the disk gas to accrete.
Although not shown in this figure, in all models, the radiation angular momentum flux ${R}^r_\phi$ contributes little to the NS's spin-up.
The reason for it is that $u^\mu\sim u_{\rm R}^\mu$ and $\rho\gg\bar{E}$ hold inside the super-Eddington accretion disk.
These relations lead to ${T_{\rm MA}}^r_\phi/R^r_\phi\sim\rho/\bar{E}\gg1$ at $R\sim r_{\rm M}$.

\begin{figure*}[tb]
\centering
\includegraphics[width=\linewidth]{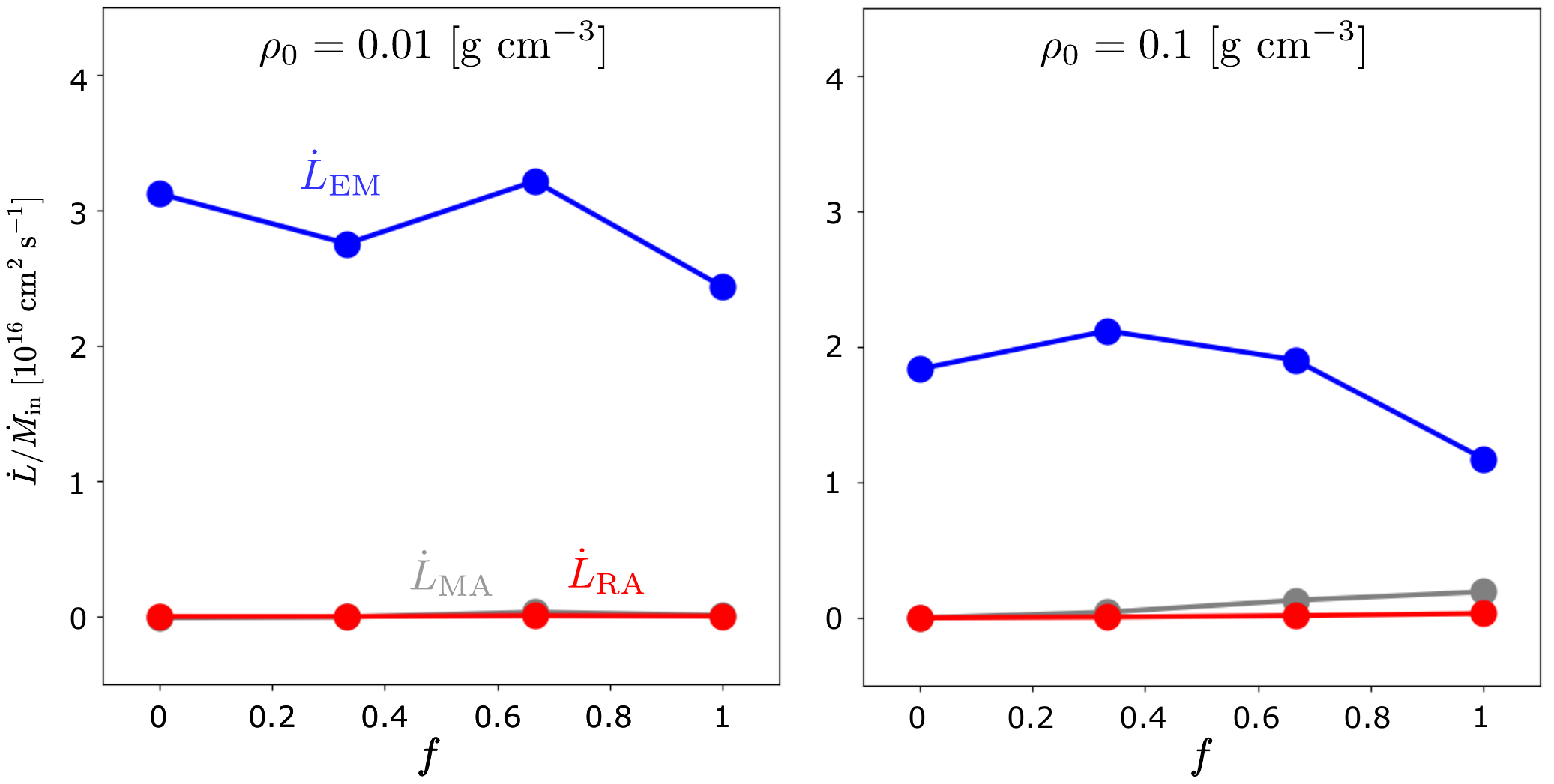}
\caption{The angular momentum flux transferred to the NS normalized by the mass accretion rate (see the text).
\label{fig:figure6}}
\end{figure*}

It is also evident from Figure \ref{fig:figure6} that the spin-up of the NS is primarily induced by the angular momentum flux of the electromagnetic field.
We plot the angular momentum fluxes of the electromagnetic field $\dot{L}_{\rm EM}$ (blue plots), of the accreting matter $\dot{L}_{\rm MA}$ (grey plots), and of the radiation $\dot{L}_{\rm RA}$ (red plots) normalized by $\dot{M}_{\rm in}$. 
They are defined as
\begin{eqnarray}
    \dot{L}_{\rm MA}&=&-2\pi\int {T_{\rm MA}}^r_\phi\sqrt{-g}d\theta,
    \label{eq:LdotMA}\\
    \dot{L}_{\rm EM}&=&-2\pi\int {T_{\rm EM}}^r_\phi\sqrt{-g}d\theta,
    \label{eq:LdotEM}\\
    \dot{L}_{\rm RA}&=&-2\pi\int R^r_\phi\sqrt{-g}d\theta.
    \label{eq:LdotRA}
\end{eqnarray}
In all models, $\dot{L}_{\rm EM}/\dot{M}_{\rm in}$ is dominant over the other fluxes.

Due to the conversion of gas angular momentum into that of the electromagnetic field near $r\sim r_{\rm M}$, the specific angular momentum of the gas there approximately coincides with $\dot{L}_{\rm EM}/\dot{M}_{\rm in}$.
The specific angular momentum at $r=r_{\rm M}$ is analytically evaluated as
\begin{eqnarray}
    l_{\rm dip}
    =1.9\times10^{16}[{\rm cm^2~s^{-1}}]
    \left(\frac{\alpha}{0.1}\right)^{1/7}
    \left(\frac{\dot{M}_{\rm in}}{100\dot{M}_{\rm Edd}}\right)^{-1/7}
    \left(\frac{B_{\rm dip}}{10^{10}~{\rm G}}\right)^{2/7}
    \left(\frac{M_{\rm NS}}{1.4M_\odot}\right)^{2/7}
    \left(\frac{r_{\rm NS}}{10~{\rm km}}\right)^{6/7}.
    \label{eq:lK_dip}
\end{eqnarray}
for $f=0$ and,
\begin{eqnarray}
    l_{\rm qua}
    =1.7\times10^{16}[{\rm cm^2~s^{-1}}]
    \left(\frac{\alpha}{0.1}\right)^{1/11}
    \left(\frac{\dot{M}_{\rm in}}{100\dot{M}_{\rm Edd}}\right)^{-1/11}
    \left(\frac{B_{\rm qua}}{10^{10}~{\rm G}}\right)^{2/11}
    \left(\frac{M_{\rm NS}}{1.4M_\odot}\right)^{4/11}
    \left(\frac{r_{\rm NS}}{10~{\rm km}}\right)^{8/11}.
    \label{eq:lK_qua}
\end{eqnarray}
for $f=1$, by calculating $r_{\rm M}$ from the balance between the radiation pressure of the self-similar solution of the slim disk \citep{Watarai1999} and the magnetic pressure arising from the NS's magnetic field \citep{Takahashi2017}.
Here, $\alpha$ denotes the viscous parameter.
By applying $\alpha=0.1$, $M_{\rm NS}=1.4M_\odot$, 
and $r_{\rm NS}=10{\rm km}$,
we obtain $l_{\rm dip}\sim2.2\times10^{16}~{\rm cm^2~s^{-1}}$ 
($\sim 3.1\times10^{16}~{\rm cm^2~s^{-1}}$) 
for model {\tt\string D\_d01} ({\tt\string D\_d001}) 
because of $B_{\rm dip}=4\times10^{10}~{\rm G}$
and $\dot{M}_{\rm in}=530\dot{M}_{\rm Edd}$ ($56\dot{M}_{\rm Edd}$).
Similarly, by setting $B_{\rm qua}=4\times10^{10}~{\rm G}$ and $\dot{M}_{\rm in}=390\dot{M}_{\rm Edd}$ ($73\dot{M}_{\rm Edd}$), $l_{\rm qua}$ of model {\tt\string Q\_d01} (model {\tt\string Q\_d001}) is calculated as $\sim2.1\times10^{16}~{\rm cm^2~s^{-1}}$ ($\sim 2.2\times10^{16}~{\rm cm^2~s^{-1}}$).
These $l_{\rm dip}$ and $l_{\rm qua}$ are consistent with $\dot{L}_{\rm EM}/\dot{M}_{\rm in}$ obtained by the present simulations.
The higher $\dot{L}_{\rm EM}/\dot{M}_{\rm in}$ in the models of $\rho_0=0.01$ than in the models with $\rho_0=0.1$ is due to the lower mass accretion rate leading to a larger $r_{\rm M}$.

Here, we note that in the models adopted in the present study, $r_{\rm M}$ does not depend significantly on $f$. 
Therefore, $\dot{L}_{\rm EM}/\dot{M}_{\rm in}$ is not so sensitive to $f$.
Simulations with a large $B_{\rm tot}$ and a small $\dot{M}_{\rm in}$, where $r_{\rm M}$ clearly depend on the magnetic field configuration, would clarify the dependence of $\dot{L}_{\rm EM}/\dot{M}_{\rm in}$ on $f$.

\subsection{Radiation flux\label{sec:radiation_flux}}

\begin{figure*}[tb]
\centering
\includegraphics[width=\linewidth]{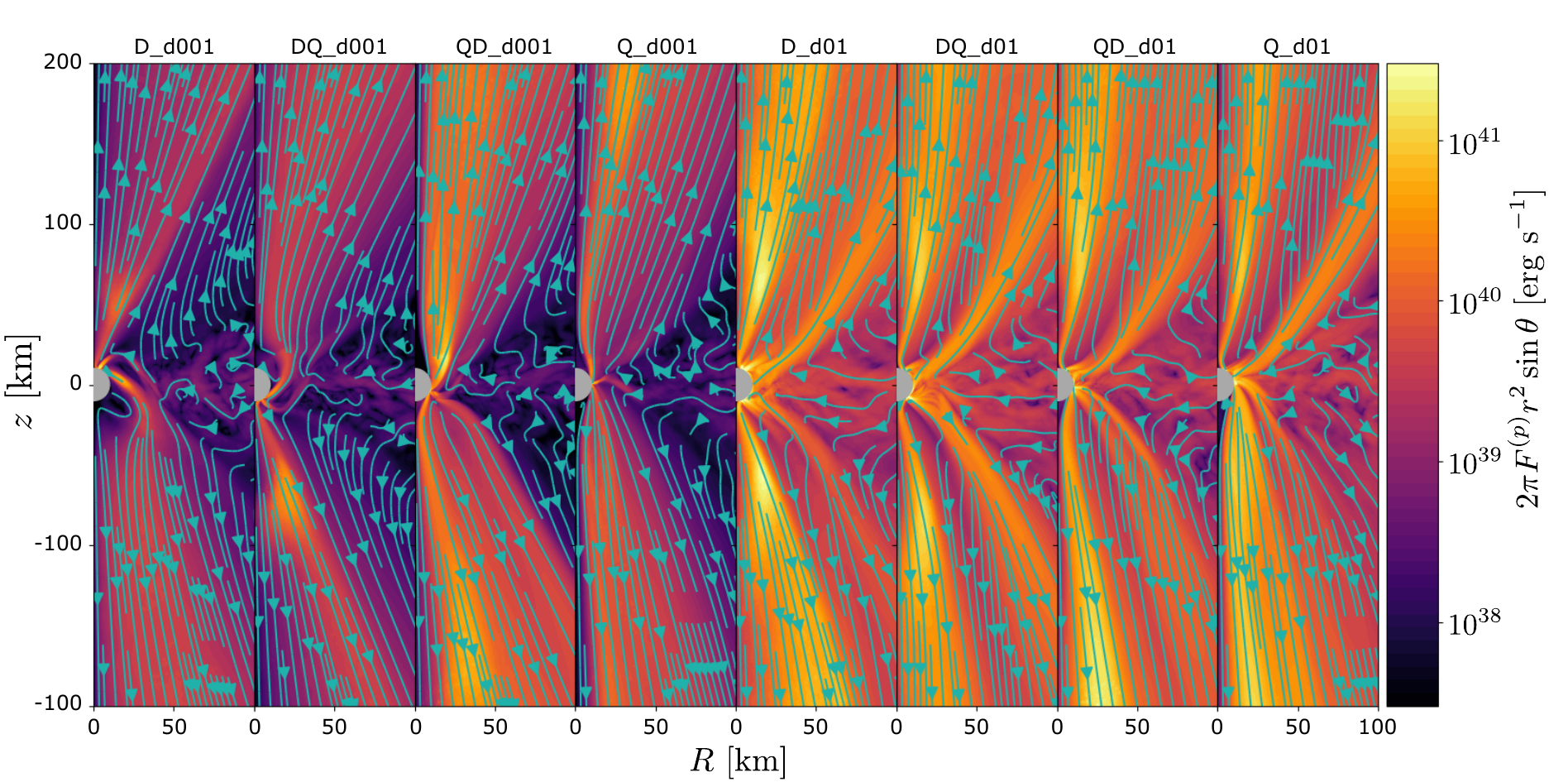}
\caption{Color maps of the norm of the radiation flux in the poloidal direction. Vectors are the radiation flux vectors.
\label{fig:figure7}}
\end{figure*}

In Figure \ref{fig:figure7}, we present the dependence of the radiation flux on $f$ and $\rho_0$.
The color shows the norm of the radiation flux in the poloidal direction $2\pi F^{(p)}r^2\sin\theta$, where $F^{(p)}=\sqrt{(R_{(t)}^{(r)})^2+(R_{(t)}^{(\theta)})^2}$.
Vectors are the radiation flux vectors in the poloidal direction measured in the static observer's frame.
It is evident that a powerful radiation flux emanates from the base of the accretion column or belt.
Although not shown in this figure, in all models, $-2\pi r^2\sin\theta\times R_t^r$ measured at the outer boundary has a peak near the rotation axis ($\theta<30^\circ,150^\circ<\theta$).
This means that the radiation from the accretion column and belt is collimated by the accretion flows and/or outflows \citep{King2017,Takahashi2017,Abarca2021,Inoue2023}.

In the following paragraphs, we explain the reason for $L_{\rm rad}^{\rm upper}\sim L_{\rm rad}^{\rm lower}$ (see, Table \ref{tab:table2}).
In models {\tt\string D\_d001}, {\tt\string DQ\_d001}, and {\tt\string QD\_001}, the same amount of the radiation energy is radiated from the accretion column towards the upper and lower hemispheres (see, Appendix \ref{sec:radiation_from_ac_ab}).
For example, in model {\tt\string D\_d001}, the radiation flux along the z-axis intensifies due to the radiation from the accretion column base towards $\theta=0^\circ$ \citep[sometime called polar beam see, e.g.,][]{Trumper2023,Kobayashi2023}.
In addition, the radiation from the accretion column base towards the equator also creates the enhanced radiation flux in the direction of lower right for $r\lesssim40~{\rm km}$.
The direction of the enhanced flux gradually changes, resulting in the radiation flux along the axis of $\theta=180^\circ$ for $r\gtrsim100~{\rm km}$.
In this way, despite the formation of the accretion column at only one pole, 
$L_{\rm rad}^{\rm upper}\sim L_{\rm rad}^{\rm lower}$ is achieved.
The resulting distribution of the radiation flux in models {\tt\string DQ\_d001} and {\tt\string QD\_d001} has almost the same trend as in model {\tt\string D\_d001}.
The bending of the radiation flux direction would be attributed to scattering by electrons in the accretion flow.
This can be inferred from the fact that the direction of the radiation energy transport is more curved than would be expected from geodesics.

In models {\tt\string D\_d01} and {\tt\string DQ\_d01}, the radiation flux exhibits symmetry with respect to the equator due to the nearly equatorial symmetry of the accretion flow.
Thus, $L_{\rm rad}^{\rm upper}\sim L_{\rm rad}^{\rm lower}$ holds.
The same amount of the radiation energy is radiated from the accretion belt located at the equator towards the upper and lower hemisphere in models {\tt\string Q\_d001}, {\tt\string QD\_d01}, and {\tt\string Q\_d01}.
This results in $L_{\rm rad}^{\rm upper}\sim L_{\rm rad}^{\rm lower}$.
Although the accretion column also exists in models {\tt\string QD\_d01} and {\tt\string Q\_d01}, the radiation from the accretion belt is more powerful than that from the accretion column.
In addition, the outward radiation flux prominent at $\theta\sim45^\circ$ and $135^\circ$ appears in models of $\rho_0=0.1~{\rm g~cm^{-3}}$.
This originates from the radiation from the accretion column and/or belt towards the less dense region ($10~{\rm km}\lesssim r \lesssim 20~{\rm km}$, $45^\circ\lesssim\theta\lesssim135^\circ$ for models {\tt\string D\_d01} and {\tt\string DQ\_d01}, $10~{\rm km}\lesssim r\lesssim 12~{\rm km}$, $45^\circ\lesssim\theta\lesssim60^\circ$ for model {\tt\string QD\_d01}, $10~{\rm km}\lesssim r\lesssim12~{\rm km}$, $45^\circ\lesssim\theta\lesssim60^\circ$ for model {\tt\string Q\_d01}).

{\section{Discussion} \label{sec:discussion}}

\subsection{The magnetic field of the NS in Swift J0243.6+6124}

In this subsection, we apply our model to ULXPs Swift J0243.6+6124 and restrict the magnetic field strength of the NS. 
To summarize, the magnetic field strength $2\times10^{13}~{\rm G}$ inferred from the CRSF observation would be originated from the quadrupole magnetic field. 
The dipole magnetic field strength $B_{\rm dip}$ would be less than $4\times10^{12}~{\rm G}$. 
These results do not contradict with the other observations, such as the thermal emission, pulse period, and spin-up rate.

We firstly discuss the case of $B_{\rm qua}\le B_{\rm dip}$.
Then, the dipole field is prominent in whole region, so that the quadrupole magnetic field would not affect the dynamics of the accretion flow.
According to the discussion in section 3.3 in \citet{Inoue2023}, $B_{\rm dip}$ is estimated from three conditions of $r_{\rm M}<r_{\rm sph}$, $130\dot{M}_{\rm Edd}<\dot{M}_{\rm in}<1200\dot{M}_{\rm Edd}$, and  $r_{\rm NS}<r_{\rm M}<r_{\rm co}$.
Here, $r_{\rm sph}=(3/2)(\dot{M}_{\rm in}/\dot{M}_{\rm Edd})r_{\rm g}$ is the spherization radius \citep{Shakura1973,Poutanen2007}, and $r_{\rm co}=[GMP^2/(2\pi)^2]^{1/3}$ is the corotation radius for the NS rotation.
When $r_{\rm M}<r_{\rm sph}$, the outflows driven by the radiation force are launched from the accretion disk.
Such outflows can explain the thermal emission observed in Swift J0243.6+6124 \citep{Tao2019} when $130\dot{M}_{\rm Edd}<\dot{M}_{\rm in}<1200\dot{M}_{\rm Edd}$.
The condition of $r_{\rm M}<r_{\rm co}$ is required for the gas to accrete without being inhibited by the centrifugal force caused by the rotation of the magnetosphere \citep{Illarionov1975}.
On the other hand, $r_{\rm M}>r_{\rm NS}$ is needed if the observed pulse originates from the accretion flows inside the magnetosphere.
In the observations of Swift J0243.6+6124, the thermal emission is detected in two observations (Obs.1 and 2), while is not detected in Obs. 3.
The pulse period $P$ is almost constant at $9.8~{\rm s}$, and the spin-up rate $\dot{P}$ is reported to be $\dot{P}=-2.22\times10^{-8}~{\rm s~s^{-1}}$ (Obs.1), $\dot{P}=-1.75\times10^{-8}~{\rm s s^{-1}}$ (Obs.2), and $\dot{P}=-6.8\times10^{-9}~{\rm s~s^{-1}}$ (Obs.3) \citep{Doroshenko2018,Chen2021}.
Applying $P$ and $\dot{P}$ in Obs.1-3 to the three conditions noted above, we get $3\times10^{11}~{\rm G}<B_{\rm dip}<4\times10^{12}~{\rm G}$.
Since $2\times10^{13}~{\rm G}$ is not included within this range, the observation of the CRSF can not be explained.
Thus, $B_{\rm qua}\le B_{\rm dip}$ is ruled out.

When $B_{\rm qua}>B_{\rm dip}$, the CRSF would originate from the quadrupole magnetic field (i.e., $B_{\rm qua}=2\times10^{13}~{\rm G}$).
In the following, we estimate the allowed range of $B_{\rm dip}$.
Firstly, we discuss the range of $B_{\rm dip}$ focusing only on Obs.1.
When $p_{\rm dip}(r_{\rm M})\ge p_{\rm qua}(r_{\rm M})$ is satisfied, dipolar accretion flows occur (see Figure \ref{fig:figure3}). 
Using equations (24) and (27) from \citet{Inoue2023}, this inequality can be rewritten as $B_{\rm dip}\ge B_{\rm dip}^{\rm trans}$.
Here, $B_{\rm dip}^{\rm trans}$ is defined as
\begin{eqnarray}
    B_{\rm dip}^{\rm trans}
    &=&
    8.6\times10^{11}~{[\rm G]}
    \left(\frac{\alpha}{0.1}\right)^{-1/5}
    \left(\frac{B_{\rm qua}}{2\times10^{13}~{\rm G}}\right)^{3/5}
    \left(\frac{M_{\rm NS}}{1.4M_\odot}\right)^{1/5}\nonumber\\
    &\times&
    \left(\frac{r_{\rm NS}}{10~{\rm km}}\right)^{-1/5}
    \left(\frac{P}{9.8~{\rm s}}\right)^{-2/5}
    \left(\frac{\dot{P}}{-10^{-8}~{\rm s~s^{-1}}}\right)^{1/5}.
    \label{eq:Btrans}
\end{eqnarray}
By substituting the observed values of $B_{\rm qua}$, $\dot{P}$, and $P$ into equation (\ref{eq:Btrans}), we find that $B_{\rm dip}^{\rm trans} = 10^{12}~{\rm G}$.
Thus, the condition of $B_{\rm dip}\ge10^{12}~{\rm G}$ is obtained.
In addition to this, the condition of $2\times 10^{10}~{\rm G}<B_{\rm dip}<5\times10^{12}~{\rm G}$ is also derived \citep[see,][for detail]{Inoue2023}.
Combining both conditions, if $B_{\rm dip}$ is within the range of $10^{12}~{\rm G}\le B_{\rm dip}<5\times10^{12}~{\rm G}$, the observations in Obs.1 can be explained by dipolar accretion flows.

On the other hand, when $p_{\rm dip}(r_{\rm M})< p_{\rm qua}(r_{\rm M})$ ($B_{\rm dip}< B_{\rm dip}^{\rm trans}=10^{12}~{\rm G}$), quadrupolar accretion flows emerge. 
In this case, the conditions $r_{\rm M} < r_{\rm sph}$, $130\dot{M}_{\rm Edd}<\dot{M}_{\rm in}<1200\dot{M}_{\rm Edd}$, and $r_{\rm NS}<r_{\rm M}<r_{\rm co}$ are all satisfied. 
First, regarding $\dot{M}_{\rm in}$, the spin-up rate is calculated using the equation $\dot{P}=\dot{M}_{\rm in}l_{\rm qua}(r_{\rm M})P/(M_{\rm NS}l_{\rm NS})$ \citep{Shapiro1983}: 
\begin{eqnarray}
    \dot{P}
    &=&
    -8.9\times10^{-10}[{\rm s~s^{-1}}]
    \left(\frac{\alpha}{0.1}\right)^{1/11}
    \left(\frac{\dot{M}_{\rm in}}{10\dot{M}_{\rm Edd}}\right)^{10/11}
    \left(\frac{B_{\rm qua}}{2\times10^{13}~{\rm G}}\right)^{2/11}\nonumber\\
    &\times&
    \left(\frac{r_{\rm NS}}{10^6~{\rm cm}}\right)^{-14/11}
    \left(\frac{M_{\rm NS}}{1.4M_\odot}\right)^{4/11}
    \left(\frac{P}{9.8~{\rm s}}\right)^2,
    \label{eq:spin-up_qua}
\end{eqnarray}
where $l_{\rm NS}$ is the specific angular momentum of the NS.
The reason $l_{\rm qua}$ is used here is that the quadrupole magnetic field is dominant at $R=r_{\rm M}$. 
From the observed values of $B_{\rm qua}$, $\dot{P}$, and $P$, $\dot{M}_{\rm in}=340\dot{M}_{\rm Edd}$ is obtained using equation (\ref{eq:spin-up_qua}), which meets the range of requirements.
Moreover, the magnetospheric radius obtained from $p_{\rm rad}=p_{\rm qua}$ is $180~{\rm km}$, and since $r_{\rm sph}=1100~{\rm km}$, the first inequality ($r_{\rm M}<r_{\rm sph}$) is satisfied.
Furthermore, because $r_{\rm co}=7700~{\rm km}$, the third inequality, $r_{\rm NS}<r_{\rm M}<r_{\rm co}$, is also satisfied.

As a result, combining both the cases of dipolar and quadrupolar accretion flows, the range of $B_{\rm dip}$ consistent with the observation of Obs.1 is $B_{\rm dip}<5\times10^{12}~{\rm G}$.
Using the same manner explained above, the dipole magnetic field strength is restricted to $B_{\rm dip}<4\times10^{12}~{\rm G}$ and $B_{\rm dip}<10^{14}~{\rm G}$ for Obs. 2 and 3, respectively.
In Table \ref{tab:table3}, we summarize the allowed range of $B_{\rm dip}$ separately for each observations (see the third row).
The ranges of $B_{\rm dip}$ for the dipolar and quadrupolar accretion flow cases are also presented individually in the first and second rows, respectively.
Assuming that the timescale of the decay of the NS's magnetic field is sufficiently long compared to the observation period, $B_{\rm dip}<4\times10^{12}~{\rm G}$ is required to explain all three observations at the same time.

\begin{deluxetable*}{lccc}[t]
\tablecaption{$B_{\rm dip}$ estimated from each observations \label{tab:table3}}
\tablehead{
\colhead{Obs. Name}
&\colhead{Obs. 1}
&\colhead{Obs. 2}
&\colhead{Obs. 3}
}
\startdata
Dipolar accretion & 
$10^{12}~{\rm G}\le B_{\rm dip}<5\times10^{12}~{\rm G}$ &
$10^{12}~{\rm G}\le B_{\rm dip}<4\times10^{12}~{\rm G}$ &
$8\times10^{11}~{\rm G}\le B_{\rm dip}<10^{14}~{\rm G}$ 
\\
Quadrupolar accretion &
$B_{\rm dip}<10^{12}~{\rm G}$ &
$B_{\rm dip}<10^{12}~{\rm G}$ &
$B_{\rm dip}<8\times10^{11}~{\rm G}$ 
\\
Combined &
$B_{\rm dip}<5\times10^{12}~{\rm G}$ &
$B_{\rm dip}<4\times10^{12}~{\rm G}$ &
$B_{\rm dip}<10^{14}~{\rm G}$ 
\\
\enddata
\end{deluxetable*}

\subsection{CRSF in Swift J0243.6+6124 and M51 ULX8\label{sec:CRSF}}

Here, we demonstrate that the line width of CRSF in Swift J0243.6+6124 and M51 ULX8 can be explained by the thermal motion of electrons inside accretion flows near the NS.

In our simulations, the gas temperatures of the accretion flows near the NS are $\sim10^{8}~{\rm K}$.
Thus, we obtain the line width of $\sigma_{\rm cyc,e}=(E_{\rm cyc}/c)\sqrt{2kT/m_{\rm e}}\sim19~{\rm keV}$ and $\sigma_{\rm cyc,p}=(E_{\rm cyc}/c)\sqrt{2kT/m_{\rm p}}\sim0.44~{\rm keV}$ for $E_{\rm cyc}=146~{\rm keV}$ \citep{Kong2022}.
The resulting $\sigma_{\rm cyc,e}$ is consistent with $\sigma_{\rm cyc}$ in Swift J0243.6+6124 ($\sigma_{\rm cyc}\sim20-30~{\rm keV}$).
On the other hand, the resulting $\sigma_{\rm cyc,p}$ is smaller than the observed $\sigma_{\rm cyc}$.
A similar conclusion can be drawn for M51 ULX8. 
Assuming $E_{\rm cyc}=4.5~{\rm keV}$, $\sigma_{\rm cyc,e}$ would be $0.58~{\rm keV}$, which is consistent with the values suggested by observations ($0.1~{\rm keV}$ or $1.0~{\rm keV}$). 
However, $\sigma_{\rm cyc,p}\sim0.013~{\rm keV}$ does not match the observed values.
Thus, our simulations suggest that the CRSF in Swift J0234.6+6124 and M81 ULX8 originates from the resonant scattering of electrons.
However, it should be noted that if the gas temperature depends on the magnetic field strength of the NS, deviations from our results may occur, potentially altering the derived $\sigma_{\rm cyc,e}$ and $\sigma_{\rm cyc,p}$ values.
Also, the optically thick accretion flows around the NS (i.e., the optically thick accretion curtain; see Figure \ref{fig:figure3}) may obscure and thus lead to the disappearance of CRSFs from the observed spectrum \citep{Mushtukov2017}.
Post-processing radiative transfer simulations are important to investigate the CRSF in detail.

\subsection{Future issues}

Simulations of the NS with strong magnetic fields are needed to check the robustness of our conclusion.
In this study, due to the numerical reason, we consider the NS with a magnetic field strength of $4\times10^{10}~{\rm G}$, and simulation results are extrapolated to the stronger magnetic field using the analytical solution of the spin-up rate.
Whether equations (\ref{eq:lK_dip}) and (\ref{eq:lK_qua}) hold for the strong magnetic field case should be confirmed directly by numerical simulations. 
Such simulations would also enable us to obtain the dependence of the spin-up rate on $f$ (see Section \ref{sec:angular_momentum_transfer}).
Recently, numerical methods that can solve the basic equations of GR-MHD stably in the high-magnetized region have been proposed \citep{Parfrey2017,Parfrey2023,Phillips2023,Chael2024}.
In the future, we plan to implement such methods in our codes.
In addition, simulations considering the reduced electron scattering opacity in the strong magnetic field are also left as important future work \citep{Sheng2023}.
When the magnetic field strength exceeds $10^{13}~{\rm G}$, the electron scattering cross section becomes smaller than the Thomson one.
In this case, the radiation flux emergent from the side wall of the column would intensify, leading to a lower gas temperature inside the accretion column \citep{Mushtukov2015b}.
This results in the line width of the CRSF narrower than estimated in Section \ref{sec:CRSF}.

Although we employ the M1 scheme in the present work, this scheme provides accurate radiation fields in optically thick regions, but when the system is optically thin and the radiation field is anisotropic, the M1 scheme can produce unphysical solutions \citep[see e.g.,][]{Asahina2020}. 
The accretion disk and outflows in our model are highly optically thick, so the adopting M1 closure is a good approximation to describe the radiation field. This approximation is, however, not appropriate near the rotation axis and in the low-density region within the magnetosphere. 
Therefore, to solve the structures near the neutron star more accurately, simulations in which the radiative transfer equation is directly solved are needed. 
Such simulations are left as very important future work \citep[see e.g.,][]{Jiang2014,Ohsuga2016,Asahina2020,Zhang2022,Asahina2022}.

Three-dimensional simulations of the super-Eddington accretion flows onto a rotating NS should be further investigated.
In this study, we perform two-dimensional simulations of a non-rotating NS with its magnetic axis aligned with the disk's rotation axis.
However, past numerical simulations showed that nonaxisymmetric modes are developed within the magnetosphere due to the magnetic Rayleigh–Taylor instability \citep{Kulkarni2008,Takasao2022,Parfrey2023,Das2024,Murguia2024,Zhu2024}.
In this case, the axisymmetric accretion columns would not form, indicating the absence of the polar beam.
Furthermore, pulsations require that the NS rotates, and the magnetic axis is misaligned with the NS's rotation axis.
Thus, we should take the nonaxisymmetric structure into account with three-dimensional simulations.
In addition, the spin-down torque was detected in two ULXPs, M82 X-2 and Swift J0243.6+6124 \citep{Bachetti2022,Liu2024,Karaferias2023}.
This detection indicates that the NSs in these ULXPs are close to spin equilibrium.
In this case, the centrifugal force due to the rotation of the NS's magnetosphere would affect the resultant spin-up rate \citep{Karaferias2023} and the estimation of the magnetic field strength.

We plan to conduct post-processing radiative transfer simulations to obtain realistic radiation spectra.
Our simulations show that the observed radition spectra can be affected by Comptonization (see Section \ref{sec:accretion_flows}).
The post-processing radiative transfer simulations let us estimate the radiation spectra including Comptonization.
If we take the anistropic electron scattering into account \citep[see, e.g.,][]{Herold1982}, the appearance of the CRSF can also be confirmed (see, Section \ref{sec:CRSF}).
Additionally, these simulations might enable us to distinguish between the observational features of dipolar and quadrupolar accretion flows.

The investigation of the super-Eddington accretion flows, whose mass accretion rate gradually changes over time, is left as an important future work.
In the present study, we examine the quasi-steady state of the super-Eddington accretion flows.
However, in the observations of Swift J0243.6+6124, the observed luminosity changes in the wide range \citep[see, e.g.,][]{Wilson2018}, which indicates that the mass accretion rate at the NS also changes.
As the mass accretion rate increases, the resulting spin-up rate would also increase over time.
To examine a more realistic situation, simulations at which the mass accretion rate changes over time are needed.

The large-scale structure of the accretion flow requires further investigation. 
In our simulations, the net mass accretion rate is almost constant (inflow-outflow equilibrium) within $r\lesssim 100~{\rm km}$, and the quasi-steady outflows mainly originate from this region. 
Therefore, even if the inflow-outflow equilibrium region expands, our conclusion regarding the estimate of the NS magnetic field strength would remain unchanged. 
This is because the estimation of the NS magnetic field strength is based on the size of the photosphere in the outflow region, and while the expansion of the equilibrium region may lead to additional outflows, these are expected to flow primarily along the equatorial plane \citep[][]{Kitaki2021,Yoshioka2022}.
Consequently, the photosphere size would not change significantly, except near the equatorial plane. 
To accurately compute the large-scale accretion flows, long-term simulations with the initial torus placed further out will be required \citep[see also, recent simulations by][]{Toyouchi2024}.

{\section{Conclusion} \label{sec:conclusion}}

In this study, we perform GR-RMHD simulations of super-Eddington accretion flows onto an NS with dipole and quadrupole magnetic fields.
The super-Eddington accretion disks form outside the magnetospheric radius and the optially-thick outflows are launched from the disk surface via the radiation force.
The inflows aligned with magnetic filed lines appear
within the magnetoshperic radius.
Thus, the accretion columns near the NS's magnetic poles form 
when the dipole magnetic field is more prominent than the quadrupole magnetic field.
For the case that the quadrupole magnetic field is dominant,
the gas preferentially accretes along the equator, forming the accretion belt.
In addition to the belt, the accretion column also forms
when the mass accretion rate is high.
In both cases, the angular momentum flux of the disk gas is converted into that of the electromagnetic field through the interaction between the gas and the NS's magnetic field lines
around the magnetospheric radius.
The inward angular momentum flux via the electromagnetic field finally causes the spin-up of the NS.
The radiation flux measured at the outer boundary of simulation domain is symmetric with respect to the equator, even if the asymmetric accretion flows form near the NS.
The reason for this is that the radiation energy is transported to 
the opposite side across the equator.
Based on our models, the observations in Swift J0243.6+6124, such as thermal emission, spin-up rate, spin periods, and CRSF can be explained by the NS with $B_{\rm dip}\lesssim 4\times10^{12}~{\rm G}$ and $B_{\rm qua} \sim 2\times10^{13}~{\rm G}$.

\begin{acknowledgments}
We would like to thank Takumi Ogawa for useful discussion.
This work was supported by JSPS KAKENHI grant Nos. JP22KJ0368, JP24KJ0143 (A.I.), JP21H04488, JP18K03710 (K.O.), JP20H00156, JP24K00678 (H.R.T), JP18K13591, and JP23K03445 (Y.A.). A part of this research has been funded by the MEXT as "Program for Promoting Researches on the Supercomputer Fugaku" 
(Toward a unified view of the universe: from large-scale structures to  planets, JPMXP1020200109; K.O., H.R.T., Y.A., and A.I.), and by Joint Institute for Computational Fundamental Science (JICFuS; K.O.). Numerical computations were performed with computational resources provided by the Multidisciplinary Cooperative Research Program in the Center for Computational Sciences, University of Tsukuba, Cray XC 50 at the Center for Computational Astrophysics (CfCA) of the National Astronomical Observatory of Japan (NAOJ), the FUJITSU Supercomputer PRIMEHPC FX1000 and FUJITSU Server PRIMERGY GX2570 (Wisteria/BDEC-01) at the Information Technology Center, The University of Tokyo.
\end{acknowledgments}

%






\appendix

{\section{Radiative shock}\label{sec:shock}}

\begin{figure}[tb]
\centering
\includegraphics[width=\linewidth]{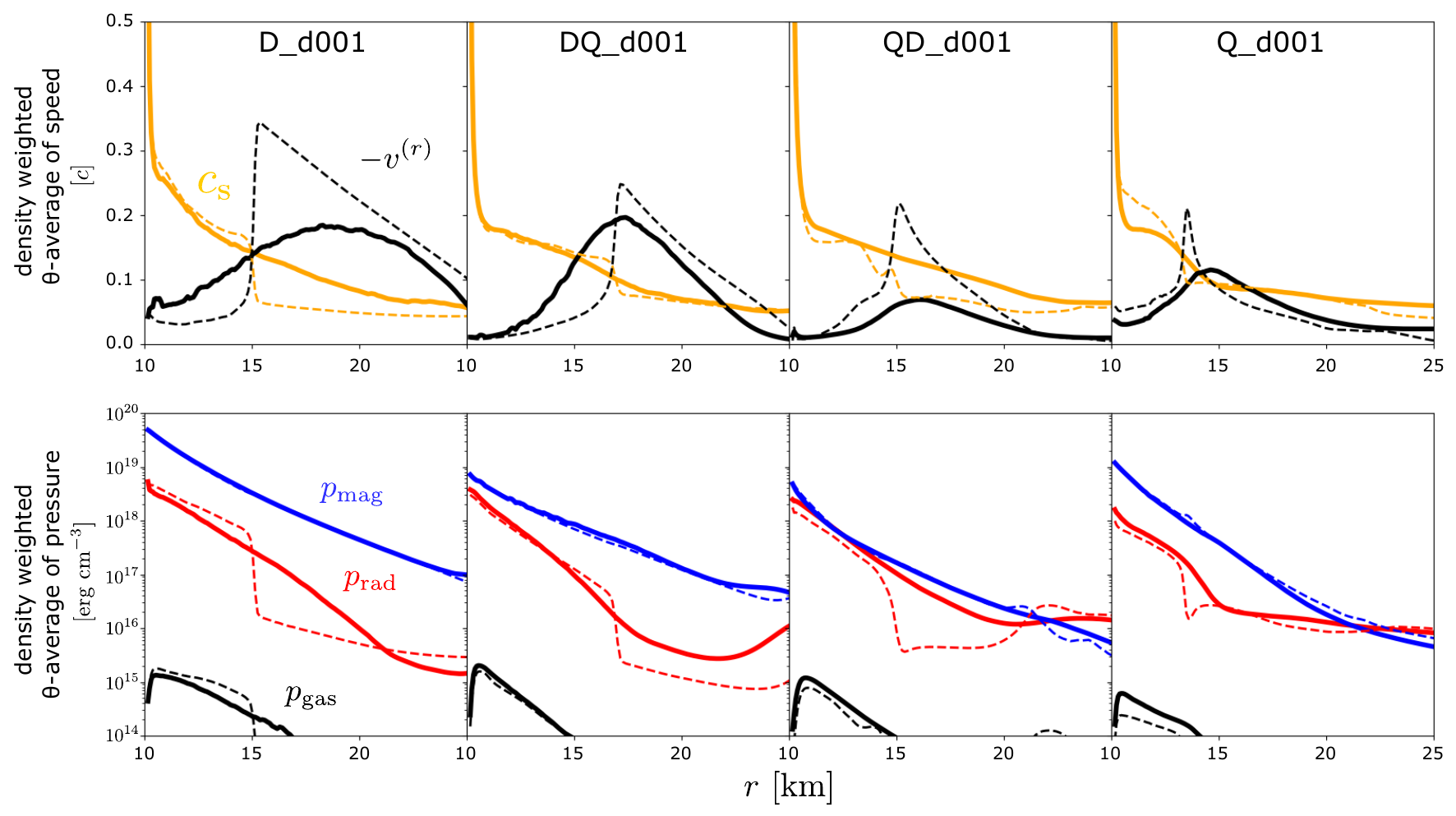}
\caption{The radial profile of the infalling velocity, sound speed, radiation pressure, gas pressure, and radiation pressure for models of $\rho=0.01~{\rm g~cm^{-3}}$.
\label{fig:figure8}}
\end{figure}

\begin{figure}[tb]
\centering
\includegraphics[width=\linewidth]{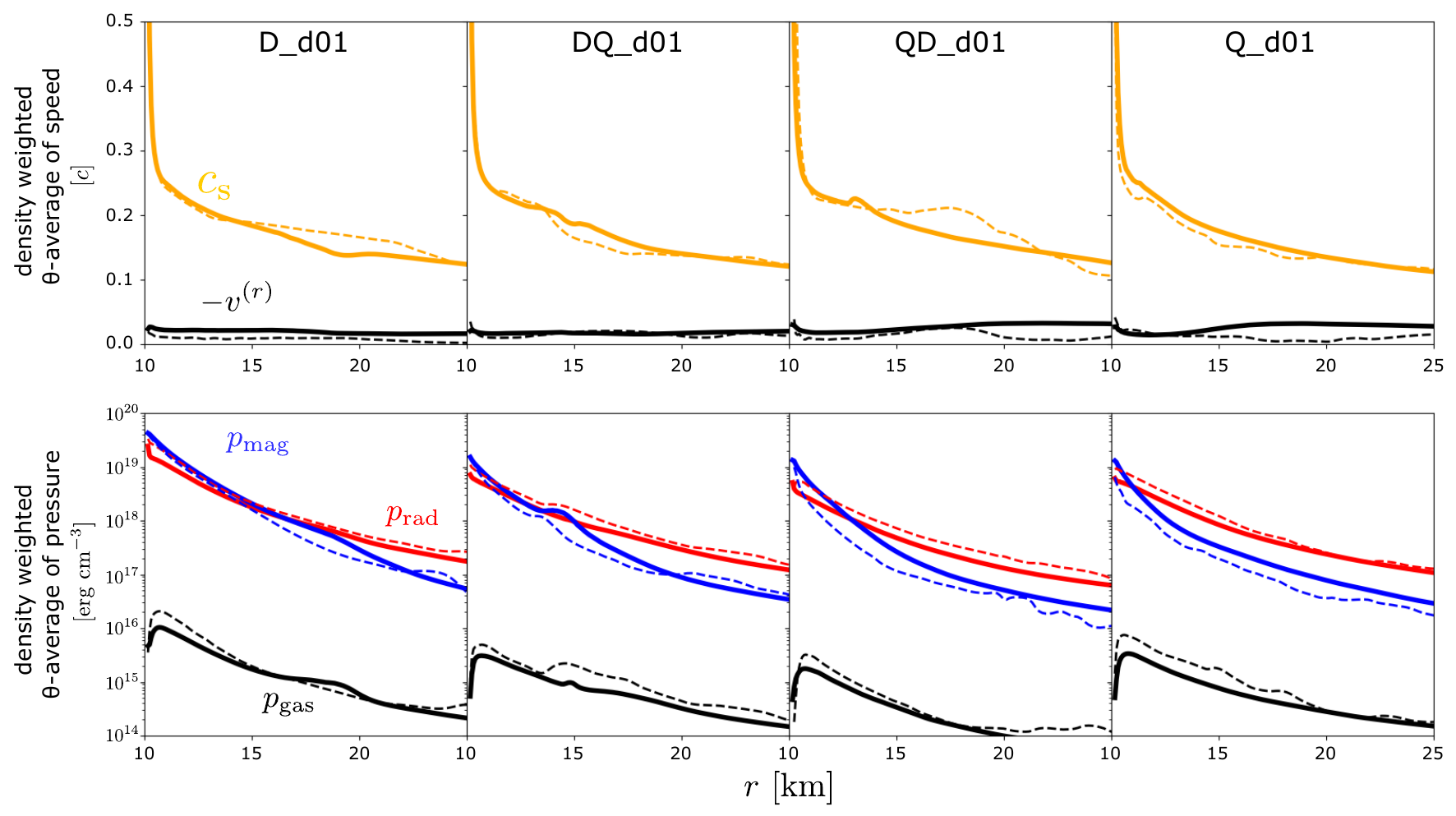}
\caption{Same as Figure \ref{fig:figure8}, but the case of $\rho=0.1~{\rm g~cm^{-3}}$.
\label{fig:figure9}}
\end{figure}

Here, we explain the radiative shock structure arising above the NS surface.
In Figures \ref{fig:figure8} and \ref{fig:figure9}, the density-weighted $\theta$-average of various quantities is plotted as a function of $r$.
The value at $t=32420t_{\rm g}$ are represented by dashed lines, while the time-averaged value is shown by solid lines.
Here, we adopt $c_{\rm s}=\sqrt{p_{\rm rad}/\rho}$ as the sound speed since $p_{\rm rad}\gg p_{\rm gas}$ inside the accretion flows.
We note that in all models, $p_{\rm mag}$ for $r\lesssim r_{\rm M}$ is greater than $p_{\rm rad}$, indicating that the accretion flows inside the magnetosphere are magnetically confined.
In the low-mass accretion rate models ($\rho_0=0.01~{\rm g~cm^{-3}}$), the radiative shock structure is clearly seen.
The discontinuity appears at $r\sim15~{\rm km}$ for model {\tt\string D\_d001}, $r\sim17~{\rm km}$ for model {\tt\string DQ\_d001}, $r\sim15~{\rm km}$ for model {\tt\string QD\_d001}, and $r\sim13~{\rm km}$ for model {\tt\string Q\_d001}.
In the post-shock region, the accreting gas is decelerated by the outward radiation force.
Thus, $-v^{(r)}$ in the post-shock region is smaller than that in pre-shock region.
On the other hand, such a shock structure cannot be seen in the high-mass accretion rate models ($\rho_0=0.1~{\rm g~cm^{-3}}$).
In these models, the infalling velocity for $r<r_{\rm M}$ is much smaller than the free-fall velocity in spite of $u_\phi$ smaller than Keplerian angular momentum.
The reason for this is that the shock surface reaches the accretion disk, and the outward radiation force keeps the infalling velocity small.
The shock structure explained above depends on time.
Therefore, the discontinuity of these quantities are smoothed out in the time-averaged profile.

{\section{Radiation from the accretion columns}\label{sec:radiation_from_ac_ab}}

\begin{figure}[tb]
\centering
\includegraphics[width=0.6\linewidth]{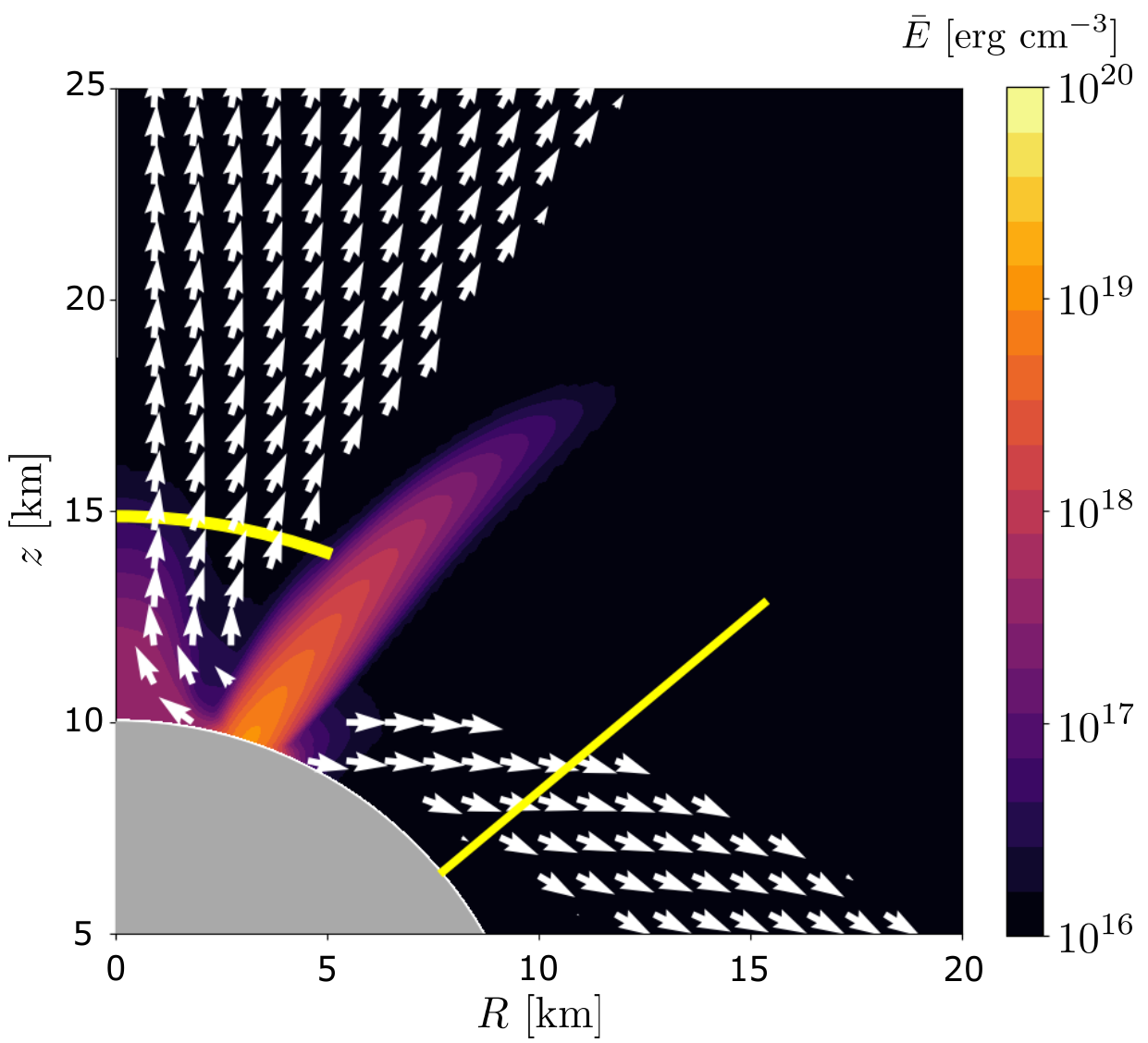}
\caption{Color map of the time-averaged radiation energy density.
Vectors are the radiation energy flux vectors in the region of $4\pi r^2F^{(p)}>10L_{\rm Edd}$.
Yellow lines represent the surface area adopted for calculating $L_{\rm rad}^1$ and $L_{\rm rad}^2$.
\label{fig:figure_MT2}}
\end{figure}

In our simulations, the same amount of the radiation energy is radiated from the accretion column towards the side with small $\theta$ and the side with large $\theta$.
Figure \ref{fig:figure_MT2} illustrates the radiation energy density near the NS with the radiation energy flux vectors $(-R_{(t)}^{(r)}, -R_{(t)}^{(\theta)})$.
Here, vectors are depicted only in the region of $4\pi r^2F^{(p)}>10L_{\rm Edd}$.
Due to the radiative shock arising above the NS surface, the radiation energy density is high inside the accretion column.
The surface area of the region where $\bar{E}$ is large has a parabolic shape \citep{Lyubarskii1988,Mushtukov2015b}.
Considerable radiation energy flux emanates from the side wall of the accretion column, and radiation towards the side with small $\theta$ changes direction near the rotation axis and goes radially.
We estimate the radiative luminosity from the side wall of the column towards the side with small $\theta$ ($L_{\rm rad}^1$) and the side with large $\theta$ ($L_{\rm rad}^2$) using
\begin{eqnarray}
    L_{\rm rad}^1
    =-2\pi\int_{0^\circ}^{30^\circ} R_{t}^{r}|_{r=15~{\rm km}}
    \sqrt{-g}d\theta,\\
    L_{\rm rad}^2
    =-2\pi\int_{10~{\rm km}}^{20~{\rm km}}R_{t}^{\theta}|_{\theta=50^\circ}
    \sqrt{-g}dr.
\end{eqnarray}
Here, the surfaces adopted for integration are drawn with yellow lines in Figure \ref{fig:figure_MT2}.
We get $L_{\rm rad}^1\sim1.0\times10^{39}~{\rm erg~s^{-1}}$ and $L_{\rm rad}^1\sim1.4\times10^{39}~{\rm erg~s^{-1}}$, and thus $L_{\rm rad}^1\sim L_{\rm rad}^2$.

\begin{figure}[tb]
\centering
\includegraphics[width=0.6\linewidth]{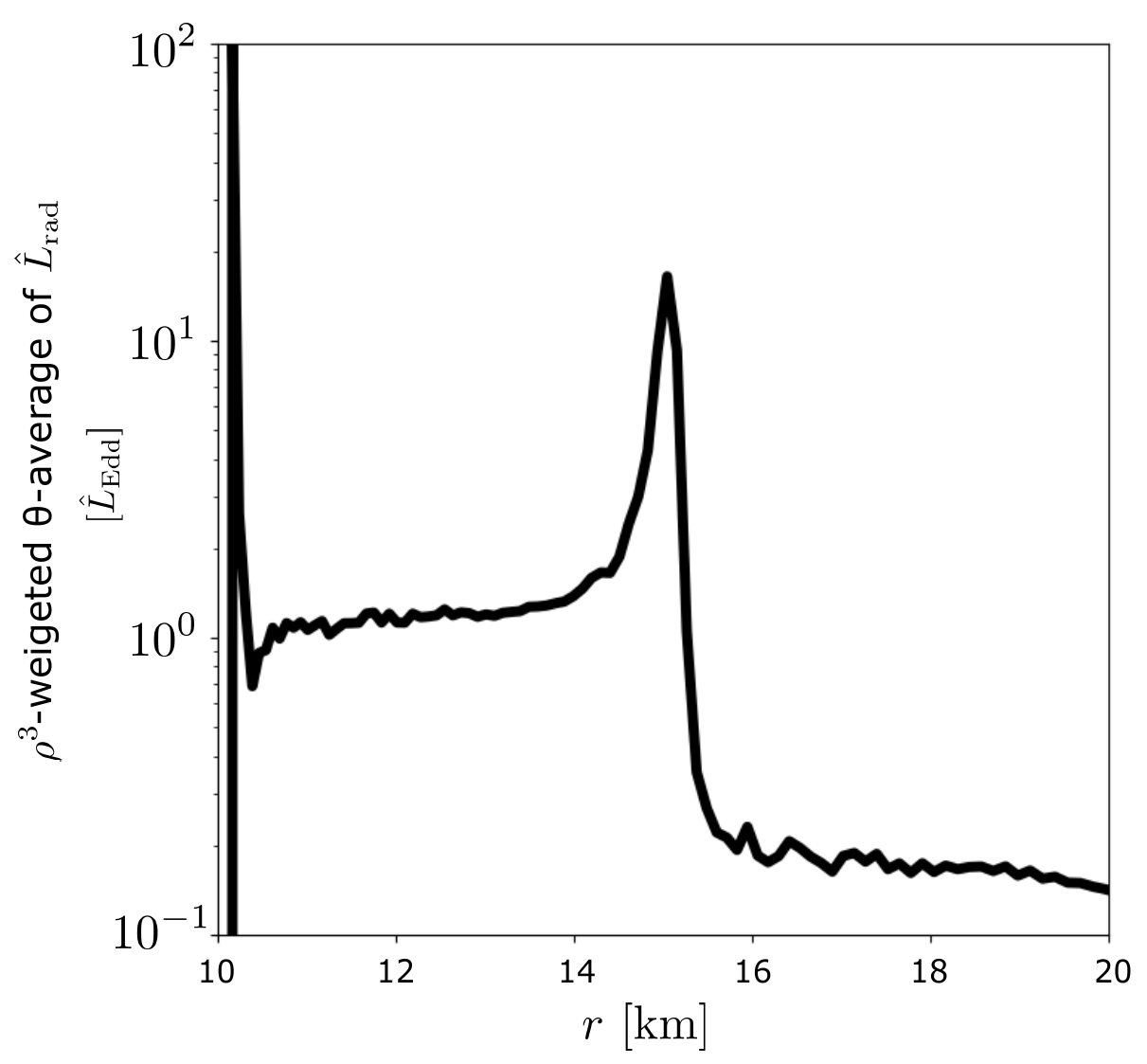}
\caption{Radial profile of the density weighted $\theta$-average of radiative luminosity in the fluid frame.
Here, we adopt the snapshot at $t=32420t_{\rm g}$.
\label{fig:figure_MT3}}
\end{figure}

We should note that although the radiative luminosity from the side wall of the column is ten times greater than $\sim L_{\rm Edd}$, the outward radiative luminosity in the fluid frame ($\hat{L}_{\rm rad}=-4\pi r^2 R^{\hat{r}}_{\hat{t}}$) inside the column is kept $\sim L_{\rm Edd}$.
Figure \ref{fig:figure_MT3} shows $\rho^3$-weighted $\theta$-average of $\hat{L}_{\rm rad}$ as a function of radius.
We can see that the resulting luminosity is almost $\sim L_{\rm Edd}$ for $11~{\rm km}\lesssim r\lesssim 14~{\rm km}$.
At $r\sim15~{\rm km}$, $\hat{L}_{\rm rad}$ has a peak due to the radiative shock.
The large $\hat{L}_{\rm rad}$ near the NS's surface would be attributed to the boundary, at which the gas is swallowed by the NS, but the energy is not swallowed by the NS.

\bibliographystyle{aasjournal}
\bibliography{main}



\end{CJK*}
\end{document}